\documentclass[a4paper,11pt]{article}

\usepackage{jheppub}

\usepackage[T1]{fontenc}
\usepackage{tikz}
\usetikzlibrary{positioning}
\usepackage[compat=1.1.0]{tikz-feynman}

\allowdisplaybreaks

\title{\boldmath The hierarchical three-body problem at $\mathcal{O}(G^2)$}

\author{Mikhail P. Solon}
\author{and Anna M. Wolz}
\affiliation{Mani L. Bhaumik Institute for Theoretical Physics, Department of Physics and Astronomy, University of California Los Angeles, Los Angeles, CA 90095, USA}

\emailAdd{solon@physics.ucla.edu}
\emailAdd{awolz@g.ucla.edu}

\abstract{
Employing techniques from scattering amplitudes and effective field theory, we model the dynamics of hierarchical triples, which are three-body systems composed of two bodies separated by a distance $r$ and a third body a distance $\rho$ away, with $r \ll \rho$. We apply the method of regions to systematically expand in the small ratio $r/\rho$ and illustrate this approach for evaluating Fourier transform integrals, which have been the bottleneck for deriving complete results in position space. In the limit where the distant third body is much heavier than the other two, we derive new analytic results in position space for the three-body conservative potential at $\mathcal{O}(G^2)$ and at leading and next-to-leading order in $r/\rho$. We also derive new results for arbitrary masses in the rest frame of the distant particle. Our results are exact in velocity, and can be used in analyses involving both bound and unbound hierarchical triples in astrophysical systems.
}

\begin{document} 
\maketitle
\flushbottom

\section{Introduction}

Systems of multiple gravitationally-interacting compact bodies are ubiquitous in astrophysics and central to many problems in stellar dynamics, exoplanet science, and gravitational waves. The case of three bodies is a classic problem in mechanics~\cite{Poincare:1890}, and of particular significance in astrophysics are so-called hierarchical triples: three-body systems composed of two bodies separated by a distance $r$, and a third body a distance $\rho$ away, with $r \ll \rho$. Hierarchical triples are common and thought to be favored by stability arguments~\cite{Tokovinin_2014,Raghavan:2010hq,Sana_2014,Ransom_2014,Naoz_2016}, with prototypical examples including the sun-earth-moon system, binary systems orbiting the supermassive black hole at the galactic center, and binary systems perturbed by a random flyby of a star. Hierarchical triples have recently been the focus of intense study in an effort to understand their impact on tidal disruption events and star ejections~\cite{Brown:2018gar,Sari:2019hot,Ginsburg:2006sx,Mockler:2023ttb}, binary populations and the evolutionary pathways of stars~\cite{DallAmico:2023neb,Thompson_2011,Hamers_2013,Antonini:2017ash,Toonen:2017yct,Fragione:2019zhm,Fragione:2019hqt,Toonen_2020,He_2017,Antonini:2012ad,Antonini:2015zsa,Stephan_2016,Stephan:2019fhf,Fragione:2020gly,Martinez_2020,Nieder:2022eti,Antognini:2015ima,Hamers_2016,Silsbee:2016djf}, the dynamics and interactions of galaxies~\cite{Kulkarni:2011kk,Mannerkoski_2021,Khan:2012dj,Koehn:2023fqu,Valtonen_1996},
and gravitational wave signals for future detectors such as LISA~\cite{Liu:2017yhr,Yunes:2010sm,Silva:2022blb,Ficarra:2023zjc,LISA:2022yao,LISAConsortiumWaveformWorkingGroup:2023arg,Bonetti:2016eif,Bonetti:2018tpf,Naoz:2012bx,Lim:2020cvm,Knee:2024mst,Hoang:2019kye,Naoz:2019sjx,Rom:2023kqm,Torres-Orjuela:2020cly,Galaviz:2010te}.

The dynamics of three-body systems is a notoriously complicated problem, both in Newtonian gravity and General Relativity (GR). In Newtonian gravity, there are only pairwise two-body interactions, and one can easily write down the Hamiltonian and equations of motion, but solutions are famously chaotic~\cite{Poincare:1890}. In GR, there are intrinsic multi-body interactions due to the non-linearity of the theory, and writing down the explicit Hamiltonian is already nontrivial. Numerical solutions are an active area of research yet are costly and require input from statistical and analytic techniques~\cite{Kol:2020sjm,Manwadkar:2021qvf,Manwadkar:2023txm,Zwart:2021qxe,Lousto:2007ji,Mukherjee_2021,Feng:2018izi}. Motivated by the prevalence of hierarchical triples in astrophysics, there have been many recent efforts using a variety of techniques to improve the modeling of effects from GR~\cite{Galaviz:2011qb,Rodriguez:2018pss,Will:2013cza,Will:2014wsa,Will:2018mcj,Will:2020tri,Jones:2022aji}.
In this paper, we build on the framework developed in Ref.~\cite{Jones:2022aji} based on tools from theoretical particle physics, such as scattering amplitudes and effective field theory (EFT), to model the dynamics of hierarchical triples. 

In the past several years, powerful tools from theoretical particle physics have been successfully applied to the two-body problem in GR, deriving many new results as a weak field expansion in powers of Newton's constant $G$, also known as the post-Minkowskian (PM) expansion\footnote{The $n$th PM order is ${\cal O}(G^n)$.} (see reviews~\cite{Adamo:2022dcm,Bjerrum-Bohr:2022blt,Buonanno:2022pgc,Kosower:2022yvp,Travaglini:2022uwo}). Remarkably, these results are exact in velocity, and may be used for modeling both bound and unbound systems. Moreover, when expanded in the limit of small velocities, also known as the post-Newtonian (PN) expansion,\footnote{The $(n+m)$th PN order is ${\cal O}(G^{n+1} v^{2m})$ for small velocities $v$.} they provide nontrivial cross-checks with results obtained via traditional approaches in GR. These developments have caught the attention of theorists involved in producing waveforms for the Ligo-Virgo-Kagra collaboration, and they found that the results can help improve the precision and efficiency of waveform models~\cite{Antonelli:2019ytb,Khalil:2022ylj,Damour:2022ybd,Rettegno:2023ghr,Buonanno:2024vkx,Buonanno:2024byg}. 

The application of these techniques to the three-body case is still at a nascent stage~\cite{Kuntz:2021ohi,Kuntz:2022onu,Loebbert:2020aos,Jones:2022aji,Ledvinka:2008tk}. The state-of-the-art result for the three-body potential in position space is at ${\cal O}(G^2)$ but expanded in the limit of small velocities (PN regime)~\cite{Loebbert:2020aos}.
One of the main obstacles for extending this to higher orders in $G$ and to all orders in velocity (PM regime) are Fourier transform integrals. In the present analysis, we focus on hierarchical triples and apply the method of regions~\cite{Beneke:1997zp} to systematically perform these integrals as an expansion in the small parameter $r/\rho \ll 1$.\footnote{While writing this paper,~\cite{Georgoudis:2023eke} described a similar method for computing the contribution from the soft region of the graviton frequency to the gravitational waveform.} In the so-called planetary limit, where the distant body is much heavier than the other two, we apply our technique to the momentum-space potential of~\cite{Jones:2022aji} and derive a new analytic result in position space at leading- and next-to-leading order (LO and NLO) in $r/\rho$. 
Additionally, we derive a new analytic result for the potential in the rest frame of the distant particle with arbitrary masses at LO and NLO in $r/\rho$.
We emphasize that the hierarchical and planetary limits are not only natural simplifications of the three-body problem but are also the relevant configuration in many astrophysical applications. Our analytic results for the conservative three-body potential at ${\cal O}(G^2)$ in position space are exact in velocity, and may be applied for hierarchical triples involving unbound trajectories, such as for modeling ejections or flybys.

The layout of the paper is as follows. In section \ref{sec:setup} we describe the physical set-up of the problem and define the hierarchical limit, and we derive the three-body ${\cal O}(G^2)$ conservative potential in momentum space and confirm the result of~\cite{Jones:2022aji}. In section \ref{sec:mor} we introduce a method of regions approach to Fourier transforms in the hierarchical limit and provide examples. 
In section \ref{sec:hampos} we calculate the position-space three-body potential in a general frame at LO and NLO in the hierarchical limit with $m_3\gg m_{1,2}$, and we additionally calculate the result for arbitrary masses in the rest frame of the distant particle. In section \ref{sec:EFTs} we formalize the method of regions analysis by matching to a sequence of EFTs. Finally, in section \ref{sec:discussion} we discuss the implications and future directions of our results.

\section{Set-up and Review}\label{sec:setup}
We are interested in the dynamics of three gravitationally-interacting compact objects with masses $m_i$ and positions $\bm{r}_i$ for $i \in \{1,2,3\}$. The distances between particles are denoted by $\bm{r}_{ij}\equiv\bm{r}_i-\bm{r}_j$, and we work in the point-particle limit where these distances are much larger than the typical sizes of the objects.

We focus on the case of hierarchical triples, composed of an ``inner binary'' and a distant perturber. To describe these, we introduce the following convenient quantities, which are used throughout the paper:
\beq\label{eq:binvars}
    \bm{r}\equiv\bm{r}_{12}\,,\qquad\bm{\rho}\equiv\bm{r}_{32}\,.
\eeq
The inner binary is composed of particles 1 and 2, separated by a distance $r$, and particle 3 perturbs the inner binary from a distance $\sim\rho$ away. 
$\bm{r}$ and $\bm{\rho}$ are the most convenient coordinates to use because they naturally show up in the Fourier transform of the three-body potential due to momentum conservation as will be shown in section~\ref{sec:mor}. The hierarchical limit is defined by
\beq
\eta\equiv\frac{r}{\rho}\ll 1\,.
\eeq
An equivalent definition of the hierarchical limit given in e.g. Ref.~\cite{Naoz_2016} is $r/R\ll 1$, where $\bm{R}$ points from the distant perturber to the Newtonian center-of-mass of the inner binary, therefore 
\beq
\bm{R}=\bm{\rho}-\frac{m_1}{m_1+m_2}\bm{r}\,.
\eeq
The system is shown in Figure~\ref{fig:figure1}.

\begin{figure}[t]\centering
\begin{tikzpicture}[roundnode/.style={circle, draw=black, very thick, minimum size=3mm}]
\filldraw[black] (0,0.8) circle [radius=0.15];
\node at (-.35,.8) {1};
\filldraw[black] (0,-0.8) circle [radius=0.15];
\node at (-.35,-.8) {2};
    \draw [<->] (0,.6) -- (0,-.6) node [midway,left] {$r$};
\filldraw[black] (10,1) circle [radius=0.15];
\node at (10.35,1) {3};
    \draw [<->,dashed] (9.8,.975) -- (0.1,0) node [midway,above] {$R$};
    \draw [<->] (9.8,.975) -- (0.2,-.8) node [midway,below] {$\rho$};
    \end{tikzpicture}
\caption{The inner binary is composed of particles 1 and 2, and has radius $\sim r$. Particle 3 perturbs the inner binary at a distance $\rho\sim R$. The hierarchical limit is defined by $r/\rho \ll 1$.}
\label{fig:figure1}
\end{figure}

To derive the gravitational three-body potential, we model the particles as scalars using the action
\beq\label{eq:QFTaction}
S=S_{\textrm{EH}} +S_{\textrm{GF}} + \int d^{d+1}x\sqrt{-g} \sum^3_{i=1} \frac12 (|\nabla_\mu\phi_i|^2-m_i^2|\phi_i|^2) \,, 
\eeq
where $S_{\textrm{EH}}$ is the usual Einstein-Hilbert term, $S_{\textrm{GF}}$ is the gauge fixing term, and we take $d=3 -2\epsilon$. We work in the mostly-minus convention.
In the present analysis, we do not consider higher-derivative interactions, such as those describing finite size effects or modifications to Einstein gravity, but they can be included straightforwardly.

We use the action in Eq.~(\ref{eq:QFTaction}) to compute scattering amplitudes perturbatively in the gravitational constant $G$, i.e., a PM expansion. We take the incoming momentum of particle $i$ to be $p_i^\mu=(E_i(\bm{p}_i),\bm{p}_i)$ and its outgoing momentum to be $(p_i-q_i)^\mu$ with $q_i^\mu=(q_i^0,\bm{q}_i)$.
Scattering amplitudes are manifestly relativistic, while the conservative potential is naturally a function of spatial 3-momenta. We therefore toggle between 3- and 4-momenta using $E_i(\bm{p}_i)=\sqrt{\bm{p}_i^2+m_i^2}$ and by using the on-shell condition to fix the energy component of the momentum transfers:
\beq\label{eq:OS}
\begin{split}
p_i^2&=(p_i-q_i)^2=m_i^2,\\
\Rightarrow q_i^0&=E_i(\bm{p}_i)-E_i(\bm{p}_i-\bm{q}_i).
\end{split}
\eeq
Note that this constraint is consistent with the potential region where $q_i^0 \sim  {{\bm p}_i \cdot {\bm q}_i \over m_i}$. In the present analysis, we consider contributions from the potential region only.

We are interested in the classical limit of scattering amplitudes, which corresponds to the limit of large charges, i.e., large masses $m_i \gg M_\textrm{Planck}$ and large angular momenta $|{\bm r}_i \times {\bm p}_j | \gg \hbar$. We implement this by scaling all graviton momenta $q_i^\mu$ by $\lambda$ and then expanding in small $\lambda$.

\subsection{Three-Body Potential in Momentum Space}\label{sec:prev}
In this subsection, we review the derivation of the contribution of intrinsic three-particle interactions in GR to the classical conservative three-body potential, calculated in momentum space and to ${\cal O}(G^2)$ in~\cite{Jones:2022aji}.

To calculate the three-body potential at ${\cal O}(G^2)$ we require the 2-to-2
scattering amplitude at ${\cal O}(G)$ and the 3-to-3 
scattering amplitude at ${\cal O}(G^2)$. These are both at tree-level and the relevant Feynman diagrams are shown in Figure~\ref{fig:alldiags}. Moreover, the classical terms scale as ${\cal O}(\lambda^{-2})$ for 2-to-2 scattering and ${\cal O}(\lambda^{-4})$ for 3-to-3 scattering. We present the results below. Further details are discussed in section 3.1 of~\cite{Jones:2022aji}.

The classical 2-to-2 scattering amplitude between particles $i$ and $j$ at ${\cal O}(G)$ in a general reference frame is

\begin{figure}[t]
\centering
 \begin{tikzpicture}[scale=.75]
        \draw[-] (0,0)--(2,0);
        \draw[-] (0,-3)--(2,-3);
        \draw[vector] (1,0)--(1,-3);
    \end{tikzpicture}
    \hspace{1.25cm}
    \begin{tikzpicture}[scale=.75]
        \draw[-] (0,0)--(2,0);
        \draw[-] (3,0)--(5,0);
        \draw[vector] (1,0)--(2.5,-1.5);
        \draw[vector] (4,0)--(2.5,-1.5);
        \draw[vector] (2.5,-1.5)--(2.5,-3);
        \draw[-] (1.5,-3)--(2.5,-3);
        \draw[-] (2.5,-3)--(3.5,-3);
    \end{tikzpicture}
    \begin{tikzpicture}[scale=.75]
        \draw[-] (0,0)--(2,0);
        \draw[-] (3,0)--(5,0);
        \draw[vector] (1,0)--(2.5,-3);
        \draw[vector] (4,0)--(2.5,-3);
        \draw[-] (1.5,-3)--(2.5,-3);
        \draw[-] (2.5,-3)--(3.5,-3);
    \end{tikzpicture}
    \begin{tikzpicture}[scale=.75]
        \draw[-] (0,0)--(2,0);
        \draw[-] (3,0)--(5,0);
        \draw[vector] (1,0)--(1,-3);
        \draw[vector] (4,0)--(4,-3);
        \draw[-] (0,-3)--(5,-3);
    \end{tikzpicture}
    \caption{All tree-level 4-point (left) and 6-point Feynman diagrams (three on the right; called Y, V, and U respectively). Straight lines are massive scalars and wavy lines are gravitons. 
    }\label{fig:alldiags}
\end{figure}

\beq
\mathcal{M}^{(2)}_{ij}=\frac{16\pi Gm_i^2m_j^2(1-2\sigma_{ij}^2)}{q_i^2}+{\cal O}(\lambda^{-1})
\eeq
with $\sigma_{ij}\equiv \frac{p_i\cdot p_j}{m_im_j}$. The classical 3-to-3 scattering amplitude between particles $i,j,$ and $k$ at ${\cal O}(G^2)$ in a general reference frame is
\beq
\mathcal{M}^{(3)}_{ijk}=\mathcal{M}^{(3)}_{\textrm{V,}ijk}+\mathcal{M}^{(3)}_{\textrm{Y,}ijk}+\mathcal{M}^{(3)}_{\textrm{U,}ijk}\,,
\eeq
with
\beq
\begin{split}
\mathcal{M}^{(3)}_{\textrm{V,}ijk}+\mathcal{M}^{(3)}_{\textrm{Y,}ijk}=\frac{-256\pi^2 G^2m_i^2m_j^2m_k^2}{q_i^2q_j^2q_k^2}\Bigg[&\frac{q_i^2(1-4\sigma_{ij}\sigma_{ik}\sigma_{jk})}{2}+\frac{(p_k\cdot q_i)^2(2\sigma_{ij}^2-1)}{m_k^2}\\
&-\frac{4(p_j\cdot q_i)(p_k\cdot q_i)\sigma_{ij}\sigma_{ik}}{m_jm_k}\Bigg]+{\cal O}(\lambda^{-3})\,,
\end{split}
\eeq
\beq
\begin{split}\label{eq:M3U}
    \mathcal{M}^{(3)}_{\textrm{U},ijk}=\frac{-256\pi^2G^2m_i^2m_j^4m_k^2}{q_i^2 q_k^2\big[(p_j+q_i)^2-m_j^2\big]}\Bigg[&(1-2\sigma_{ij}^2)(1-2\sigma_{jk}^2)-\frac{4\sigma_{jk}(1-2\sigma_{ij}^2)(p_k\cdot q_i)}{m_j m_k}\\
    & +2\Bigg(\frac{(1-2\sigma_{ij}^2)\sigma_{jk}(p_j\cdot q_k)}{m_jm_k}+\frac{(1-2\sigma_{jk}^2)\sigma_{ij}(p_j\cdot q_i)}{m_im_j}\Bigg)\Bigg]\\
    &+{\cal O}(\lambda^{-3})\,,
\end{split}
\eeq
agreeing with~\cite{Jones:2022aji}.
Note that there are ``super-classical'' terms in $\mathcal{M}^{(3)}_{\textrm{U},ijk}$ that are ${\cal O}(\lambda^{-5})$. They cannot contribute to the classical potential, which homogeneously scales as ${\cal O}(\lambda^{-4})$. They will drop out when the full theory amplitude is matched to the EFT amplitude.

The effective Hamiltonian 
\beq\label{eq:Ham}
\begin{split}
H(\{\bm{p},\bm{r}\})&=\sum_{i=1}^3E_i(\bm{p}_i)+V(\{\bm{p},\bm{r}\})
\end{split}
\eeq
is obtained by integrating out potential-mode gravitons from the full theory. The position-space potential $V(\{\bm{p},\bm{r}\})$ encodes the conservative dynamics of the system and is related to the momentum-space potential via
\beq\label{eq:Pot}
\begin{split}
V(\{\bm{p},\bm{r}\})&=\underset{{\bm{q}_1, \bm{q}_2, \bm{q}_3 }}\int e^{i\sum_{i=1}^3\bm{q}_i\cdot\bm{r}_i} \, V\big(\{\bm{p},\bm{q}\}\big) \,, \qquad
\underset{{\bm{q}_1, \bm{q}_2, \cdots, \bm{q}_N }}\int \equiv \ \prod_{i=1}^{N} \int\frac{d^d\bm{q}_i}{(2\pi)^{d}} \,.
\end{split}
\eeq
We compute this using dimensional regularization in $d=3-2\epsilon$ when necessary. We explicitly write the intrinsic two- and three-body contributions to the potential as in~\cite{Jones:2022aji}
\beq\label{eq:V3formal}
\begin{split}
V\big(\{\bm{p},\bm{q}\}\big)=\sum _{(i,j,k)\in S_3}\bigg[&\frac{1}{2}(2\pi)^{2d}\delta^{(d)}(\bm{q}_i+\bm{q}_j)\delta^{(d)}(\bm{q}_k)V_{ij}^{(2)}(\bm{p}_i,\bm{p}_j;\bm{q}_i,\bm{q}_j)\\
&+(2\pi)^{d}\delta^{(d)}(\bm{q}_i+\bm{q}_j+\bm{q}_k)V_{ijk}^{(3)}(\bm{p}_i,\bm{p}_j,\bm{p}_k;\bm{q}_i,\bm{q}_j,\bm{q}_k)\bigg]
\end{split}
\eeq
where $S_3$ is the set of distinct permutations of $(1,2,3)$. Note that $V\big(\{\bm{p},\bm{q}\}\big)$ is classical and scales homogeneously as ${\cal O}(\lambda^{-4})$.

Our goal is to compute $V^{(3)}_{ijk}\big(\{\bm{p},\bm{q}\}\big)\big|_{G^2}$, the intrinsic three-body interaction potential at ${\cal O}(G^2)$. The potential is related to the 
scattering amplitude by solving the Lippmann-Schwinger equation using the Born series~\cite{Lippmann:1950zz,Cristofoli:2019neg} order-by-order in $G$, and the result is
\beq\label{eq:LSnew}
V^{(3)}_{ijk}\big(\{\bm{p},\bm{q}\}\big)\big|_{G^2}\overset{!}{=}-M^{(3)}_{ijk}\big(\{\bm{p},\bm{q}\}\big)\big|_{G^2}-\frac{V^{(2)}_{ij}\big(\bm{p}_i,\bm{p}_j;\bm{q}_i\big)\big|_{G}V^{(2)}_{kj}\big(\bm{p}_k,\bm{p}_j+\bm{q}_i;\bm{q}_k\big)\big|_{G}}{E_j(\bm{p}_j)+q_i^0-E_j(\bm{p}_j+\bm{q}_i)}
\eeq
where $\overset{!}{=}$ denotes equality when energy is conserved between the initial and final states
\beq
\sum_{i=1}^3 E_i(\bm{p}_i)=\sum_{i=1}^3 E_i(\bm{p}_i-\bm{q}_i)
\eeq
and the EFT amplitude $M^{(3)}_{ijk}$ is related to that obtained from the full theory above by a nonrelativistic normalization
\beq
    M^{(3)}_{ijk}=\prod_{n=1}^3\frac{1}{2\sqrt{E_n(\bm{p}_n)E_n(\bm{p}_n-\bm{q}_n)}}\mathcal{M}^{(3)}_{ijk} \,.
\eeq
In~\eqref{eq:LSnew}, 3-momentum conservation is implicit in each of the $V^{(2)}\big|_{G}$ factors (the momentum transfer is the third argument) and we have written out one $(i,j,k)$ contribution, but the final answer will be the sum over all permutations of $(1,2,3)$ as usual. The second term on the RHS of~\eqref{eq:LSnew} is the so-called ``iteration term,'' depicted as an EFT diagram with vertices $V^{(2)}\big|_{G}$ in Figure~\ref{fig:V2diags}.

\begin{figure}[t]
\centering
    \begin{tikzpicture}[scale=1]
        \draw[-] (0,0)--(4,0);
        \draw[-] (0,-1)--(4,-1);
        \draw[-] (0,-2)--(4,-2);
        \filldraw[fill=gray!5, line width=1.2](1,-.5) ellipse (0.6 and .5) node {$V^{(2)}|_{\scriptscriptstyle G}$};
        \filldraw[fill=gray!5, line width=1.2](3,-1.5) ellipse (0.6 and .5) node {$V^{(2)}|_{\scriptscriptstyle G}$};
        \node at (-.2,0) {$i$};
        \node at (-.2,-1) {$j$};
        \node at (-.2,-2) {$k$};
    \end{tikzpicture}
    \caption{Connected tree-level contribution to the three-body ${\cal O}(G^2)$ potential composed of iterated two-body potentials $V^{(2)}\big|_G$.}\label{fig:V2diags}
\end{figure}

To get $V^{(2)}_{ij}\big|_{G}$ explicitly we solve the $N=2$ Lippmann-Schwinger at ${\cal O}(G)$. It has no iteration contributions, so we calculate it directly from the tree-level 2-to-2 amplitude:
\beq
V_{ij}^{(2)}\big(\bm{p}_i,\bm{p}_j;\bm{q}_i\big)\big|_{G}=\frac{4\pi G m_i^2m_j^2(2\sigma_{ij}^2-1)}{E_i E_j q_{ij}^2}\,,
\eeq
where the graviton pole is
\beq
\begin{split}
q_{ij}^2 \equiv \omega_{ij}^2-\bm{q}_i^2\,, \qquad\omega_{ij}\equiv \frac{(\bm{p}_i+\bm{p}_j)\cdot\bm{q}_i}{E_{ij}}
\end{split}
\eeq
and we used the shorthand $E_i\equiv E_i(\bm{p}_i)$ and $E_{ij}\equiv E_i+E_j$. With this form for the energy component of the graviton, the two-body potential reduces to isotropic gauge in the COM frame, where $\bm{p}_i=-\bm{p}_j \Rightarrow \omega_{ij}=0$. This choice was deemed ``generalized isotropic gauge'' in~\cite{Jones:2022aji}.

We are now ready to solve for $V^{(3)}_{ijk}\big(\{\bm{p},\bm{q}\}\big)\big|_{G^2}$. There are some features of the potential that we wish to make manifest: it should be free of matter poles that diverge as ${\bm p}_i \to 0$, and it should not have super-classical terms that scale as $\mathcal{O}\left(\lambda^{-5}\right)$, only classical terms that scale as $\mathcal{O}\left(\lambda^{-4}\right)$. Accordingly, we organize contributions to the potential as
\beq
\begin{split}
V^{(3)}_{ijk}&=V^{(3)}_{\textrm{V,}ijk}+V^{(3)}_{\textrm{Y,}ijk}+V^{(3)}_{\textrm{sub,}ijk}\,,\\
V^{(3)}_{\textrm{V,Y;}ijk}&\overset{!}{=}-\frac{\mathcal{M}^{(3)}_{\textrm{V,Y;}ijk}}{8E_i  E_j E_k }\,,\\
V^{(3)}_{\textrm{sub,}ijk}&\overset{!}{=}-\frac{\mathcal{M}^{(3)}_{\textrm{U},ijk}}{8E_i E_j E_k}-\frac{V^{(2)}_{ij}\big(\bm{p}_i,\bm{p}_j;\bm{q}_i\big)\big|_{G}V^{(2)}_{kj}\big(\bm{p}_k,\bm{p}_j+\bm{q}_i;\bm{q}_k\big)\big|_{G}}{E_j+q_i^0-E_j'}\,,
\end{split}
\eeq
where the prime denotes the shift $\bm{p}_j\rightarrow\bm{p}_j+\bm{q}_i$, so $E_j'\equiv E_j(\bm{p}_j+\bm{q}_i)$.
The terms $V^{(3)}_{\textrm{V,}ijk}$ and $V^{(3)}_{\textrm{Y,}ijk}$ are manifestly free of matter poles and purely classical. On the other hand, $V^{(3)}_{\textrm{sub,}ijk}$ contains both matter poles and super-classical terms, which are spurious and cancel as we discuss in section~\ref{sec:mat}. The final momentum-space result is 
\beq\label{eq:Vmom}
V^{(3)}(\{\bm{p},\bm{q}\}) = (2\pi)^3\delta^{(3)}(\bm{q}_1+\bm{q}_2+\bm{q}_3) \sum_{(i,j,k)\in S_3} V_{ijk}^{(3)}(\{\bm{p},\bm{q}\})
\eeq
with 
\beq
\begin{split}
    V^{(3)}_{ijk}&=\frac{32\pi^2 G^2m_i^2m_j^2m_k^2}{E_i E_j E_kq_i^2q_j^2q_k^2}\Bigg[\frac{q_i^2(1-4\sigma_{ij}\sigma_{ik}\sigma_{jk})}{2}+\frac{(p_k\cdot q_i)^2(2\sigma_{ij}^2-1)}{m_k^2}-\frac{4(p_j\cdot q_i)(p_k\cdot q_i)\sigma_{ij}\sigma_{ik}}{m_j m_k}\Bigg]\\
     &\quad-V^{(2)}_{ij}V^{(2)}_{kj}\Bigg[
    \begin{aligned}[t]&\frac{1}{2E_j}-\frac{2\omega_{kj}^2}{E_{jk}q_{kj}^2}+\frac{E_j(q_i\cdot q_k)(\omega_i+\omega_{ij})(\omega_k+\omega_{kj})}{2E_{ij}q_i^2E_{jk}q_k^2}\\
    &+\frac{E_j}{E_{jk}^2q_{kj}^2}\bigg(2\omega_i\omega_{kj}+q_i\cdot q_k\bigg(1-\frac{(\omega_k+\omega_{kj})^2}{q_k^2}\bigg)\bigg)\Bigg]\end{aligned}\\
    &\quad+16\pi G m_j m_k \sigma_{jk}V^{(2)}_{ij}\Bigg[\frac{1}{E_jq_{kj}^2}-\frac{p_k\cdot q_i}{E_kq_k^2}\Bigg(\frac{\omega_i+\omega_{ij}}{E_{ij}q_i^2}-\frac{\omega_k+\omega_{kj}}{E_{jk}q_{kj}^2}\Bigg)\Bigg]\,.
    \end{split}\label{eq:Vijk}
    \eeq
Here we used the shorthand $V_{ij}^{(2)} =V^{(2)}_{ij}\big(\bm{p}_i,\bm{p}_j;\bm{q}_i\big)\big|_{G} $ and $V_{kj}^{(2)} = V^{(2)}_{kj}\big(\bm{p}_k,\bm{p}_j;\bm{q}_k\big)\big|_{G} $, and used \eqref{eq:OS} to set $q_i^0 \approx { \bm{p}_i \cdot \bm{q}_i \over E_i} \equiv \omega_i$. This result matches Eq.~(3.15) in~\cite{Jones:2022aji}.\footnote{The minus sign in front of the $p_k\cdot q_i$ in the last line corrects a typo in~\cite{Jones:2022aji}.}

\subsection{Cancellation of Matter Poles}\label{sec:mat}
In a new derivation, we explicitly show the cancellation of spurious matter poles and super-classical terms. This derivation differs from the one presented in~\cite{Jones:2022aji} since it organizes the calculation in terms of relativistic variables, which expose the pole structure of the amplitudes and lead to more compact analytic expressions. 

The cancellation is guaranteed by the fact that the connected iteration contribution in the EFT (Figure~\ref{fig:V2diags}) and the U-type diagram in the full theory (rightmost diagram in Figure~\ref{fig:alldiags}) agree on the factorization channel given by cutting the intermediate connecting matter line. This suggests that the cancellation can be made explicit by expanding $V^{(3)}_{\textrm{sub,}ijk}$ about that matter pole. We choose the matter pole to be the full theory matter propagator and denote it as
\beq
\frac{1}{Y_{ij}}\equiv \frac{1}{(p_j+q_i)^2-m_j^2}=\frac{1}{2p_j\cdot q_i +q_i^2}\,.
\eeq
In terms of $Y_{ij}$, the EFT matter pole (including a factor of $1/(2E_j')$) in the second term of $V^{(3)}_{\textrm{sub,}ijk}$ can be written as 
\begin{align}
\frac{1}{2E_j'(E_j+q_i^0-E_j')} &=\frac{1}{Y_{ij}}+\frac{1}{4E_j^2}+{\cal O}(\lambda) \,. 
\end{align}
The graviton pole ${1 / q_{ij}^2 }$ in $V^{(2)}_{ij}\big(\bm{p}_i,\bm{p}_j;\bm{q}_i\big)\big|_{G}$ can be written in terms of the full theory graviton pole $1/q_i^2$ as
\begin{align}\label{eq:EFTleft}
{1 \over q_{ij}^2 } &= {1 \over q_i^2} + {Y_{ij} (q_i^0 + \omega_{ij} ) \over 2 E_{ij} q_i^2 q_{ij}^2} \,.
\end{align}
The graviton pole $1/q_{kj}'^2$ in $V^{(2)}_{kj}\big(\bm{p}_k,\bm{p}_j+\bm{q}_i;\bm{q}_k\big)\big|_{G}$ can be written in terms of the full theory graviton pole $1/q_k^2$ as
\beq \label{eq:EFTright}
    \frac{1}{q_{kj}'^2}\equiv \frac{1}{\Big(\frac{(\bm{p}_k+\bm{p}_j+\bm{q}_i)\cdot\bm{q}_k}{E_j'+E_k}\Big)^2-\bm{q}_k^2}=\frac{1}{\hat{q}_{kj}^2}-\frac{1}{q_{kj}^4}\frac{\omega_{kj}^2Y_{ij}}{E_jE_{jk}}+{\cal O}(\lambda^0)\,,
\eeq
where
\beq
{1 \over \hat{q}_{kj}^2 } \equiv \frac{1}{\Big(\frac{(\bm{p}_k+\bm{p}_j+\bm{q}_i)\cdot\bm{q}_k}{E_{jk}+q_i^0}\Big)^2-\bm{q}_k^2} =\frac{1}{q_k^2}-\frac{Y_{ij}(q_k^0+\hat{\omega}_{kj})}{2(E_{jk}+q_i^0)q_k^2\hat{q}_{kj}^2} \, , \qquad \hat{\omega}_{kj} \equiv \frac{(\bm{p}_k+\bm{p}_j+\bm{q}_i)\cdot\bm{q}_k}{E_{jk}+q_i^0}\,. 
\eeq
These expressions demonstrate that the matter and graviton poles in the full theory and EFT are identical when expanded to leading order in $Y_{ij}$, i.e., 
\begin{align}
\frac{1}{2E_j'(E_j+q_i^0-E_j')} \to \frac{1}{Y_{ij}} \,, \qquad 
{1 \over q_{ij}^2 } \to {1 \over q_i^2}  \,, \qquad
{1 \over q_{kj}'^2 }  \to \frac{1}{q_k^2} \, .
\end{align}
These properties are key for demonstrating the cancellation of spurious poles.

The U-graph contribution to $V^{(3)}_{\textrm{sub,}ijk}$ is then
\beq\label{eq:MEFTU}
\begin{split}
    - \frac{\mathcal{M}^{(3)}_{\textrm{U},ijk}}{8E_i E_j E_k}=\frac{32\pi^2G^2m_i^2m_j^4m_k^2}{E_iE_jE_k q_i^2 q_k^2}\frac{1}{Y_{ij}}\Bigg[&(1-2\sigma_{ij}^2)(1-2\sigma_{jk}^2)-\frac{4\sigma_{jk}(1-2\sigma_{ij}^2)(p_k\cdot q_i)}{m_j m_k}\\
    & +2\Bigg(\frac{(1-2\sigma_{ij}^2)\sigma_{jk}(p_j\cdot q_k)}{m_jm_k}+\frac{(1-2\sigma_{jk}^2)\sigma_{ij}(p_j\cdot q_i)}{m_im_j}\Bigg)\Bigg].
\end{split}
\eeq
Recall that this is just one contribution to the potential. We will sum over all distinct permutations of $(1,2,3)$. We can thus inspect symmetry properties under swapping the indices $ijk$ to anticipate cancellations among such permutations. For instance, using momentum conservation~$q_1^\mu + q_2^\mu +q_3^\mu=0$ and the on-shell condition $2p_j\cdot q_j=q_j^2$, note that
\beq\label{eq:Yid}
Y_{ij} = - Y_{kj} - 2 q_i \cdot q_k \, .
\eeq
This identity implies that, up to higher order quantum corrections, the second line of \eqref{eq:MEFTU} is anti-symmetric in $i\leftrightarrow k$ and therefore cancels when we add the $M^{(3)}_{U,kji}$ term. On the other hand, note that the first term of \eqref{eq:MEFTU} would not completely cancel since it is anti-symmetric in $i\leftrightarrow k$ only at ${\cal O}(\lambda^{-5})$ but not at ${\cal O}(\lambda^{-4})$. Considering these cancellations, we thus have
\beq\label{eq:MEFTU2}
\begin{split}
    -\frac{\mathcal{M}^{(3)}_{\textrm{U},ijk}}{8E_i E_j E_k}=\frac{32\pi^2G^2m_i^2m_j^4m_k^2 (1-2\sigma_{ij}^2)}{E_iE_jE_k q_i^2 q_k^2}\frac{1}{Y_{ij}}\Bigg[&(1-2\sigma_{jk}^2)- \frac{4\sigma_{jk}(p_k\cdot q_i)}{m_j m_k}\Bigg].
\end{split}
\eeq
The contribution to $V^{(3)}_{\textrm{sub,}ijk}$ from the iteration term is 
\beq\label{eq:Viter}
\begin{split}
    &-\frac{V^{(2)}_{ij}\big(\bm{p}_i,\bm{p}_j;\bm{q}_i\big)\big|_{G}V^{(2)}_{kj}\big(\bm{p}_k,\bm{p}_j+\bm{q}_i;\bm{q}_k\big)\big|_{G}}{E_j+q_i^0-E_j'}\\
    &\qquad =-\frac{32 \pi^2 G^2 m_i^2 m_j^4 m_k^2(1-2\sigma_{ij}^2)}{E_i E_jE_k q_{ij}^2q_{kj}'^2}\\
    &\qquad \quad \times {1 \over Y_{ij} } \Bigg[ (1-2\sigma_{jk}^2) - {4 \sigma_{jk} (p_k \cdot q_i) \over m_j m_k}   + Y_{ij} \Bigg(\frac{(1-2\sigma_{jk}^2)}{4E_j^2}+\frac{2\sigma_{jk} E_k}{m_jm_k E_j }\Bigg)  \Bigg] \,.
\end{split}
\eeq
Here we have expanded up to ${\cal O}(\lambda^{-4})$, noting the dependence of $V^{(2)}_{kj}\big(\bm{p}_k,\bm{p}_j+\bm{q}_i;\bm{q}_k\big)\big|_{G}$ on $\bm{p}_j+\bm{q}_i$. 

Note that \eqref{eq:MEFTU2} and the first two terms of \eqref{eq:Viter} would cancel up to the difference between the graviton poles in the full theory and those in the EFT given in \eqref{eq:EFTleft} and \eqref{eq:EFTright}. The remaining task is then to plug in the expressions for the graviton poles, and expand to ${\cal O}(\lambda^{-4})$. We note that, aside from the $1/Y_{ij}$ poles, we also encounter ${\cal O}(\lambda^{-5})$ superclassical pieces, which are antisymmetric in $i \leftrightarrow k$ and thus cancel upon adding the $M^{(3)}_{U,kji}$ term. Upon collecting all contributions, the final result is given in Eq.~\eqref{eq:Vmom}.

\section{Method of Regions}
\label{sec:mor}
Our goal is to obtain the position-space potential defined by Eqs.~\eqref{eq:Ham}~and~\eqref{eq:Pot}. However, the Fourier transform of \eqref{eq:Vijk} from momentum or $\bm{q}_i$-space to position or $\bm{r}_i$-space is non-trivial.
 We simplify the problem by exploiting the hierarchical limit in momentum space and using the method of regions to perform the integrals. The hierarchical limit of the three-body problem is clearly defined in position space by $r\ll \rho$, but in momentum space the limit is less obvious. Here, we describe how the hierarchical limit manifests itself in the Fourier transform of the momentum-space potential.

To start, we eliminate $\bm{q}_2$ using conservation of momentum, so the Fourier transform becomes
\beq\label{eq:FTe2}
V^{(3)}\big(\bm{p}_1,\bm{p}_2,\bm{p}_3;\bm{r},\bm{\rho}\big)=\underset{\bm{q}_1,\bm{q}_3}\int e^{i\bm{q}_1\cdot\bm{r}}e^{i\bm{q}_3\cdot\bm{\rho}}\, V^{(3)}\big(\bm{p}_1,\bm{p}_2,\bm{p}_3;\bm{q}_1,-\bm{q}_1-\bm{q}_3,\bm{q}_3\big) \, ,
\eeq
where $\bm{r}$ and $\bm{\rho}$ are defined in \eqref{eq:binvars}. Some of the integrals required for $V^{(3)}\big(\{\bm{p},\bm{r}\}\big)$ are unknown in closed form so we exploit the hierarchical limit ($r\ll\rho$) to simplify the integrand. 

In the hierarchical limit, we are interested in contributions where $|\bm{q}_3|\sim 1/\rho$. The physics at scale $r$ involves particles 1 and 2, which comprise the inner binary, and so $|\bm{q}_3|\sim 1/r$ is not relevant. We verify this in explicit examples such as those presented below. 

For the integral over $\bm{q}_1$, there are two regions that contribute:
\beq\label{eq:abscale}
\begin{split}
    &\textrm{Region a:} \qquad |\bm{q}_1|\sim \frac{1}{r}\,,\\
    &\textrm{Region b:} \qquad |\bm{q}_1|\sim \frac{1}{\rho}\,.
\end{split}
\eeq
In both regions, we use $\eta = r/\rho \ll 1$. In region a, this implies that $|\bm{q}_1|\gg |\bm{q}_3|$, and we will use this to expand integrands prior to integration. For example,
\beq
\frac{1}{(\bm{q}_1+\bm{q}_3)^2}=\frac{1}{\bm{q}_1^2}\left(1-\frac{2(\bm{q}_1\cdot \bm{q}_3)}{\bm{q}_1^2}+{\cal O}(\eta^2)\right)\,,
\eeq
which then factorizes the Fourier transforms over $\bm{q}_1$ and $\bm{q}_3$ into separate integrals. 
In region b, we have $\bm{q}_1\cdot\bm{r} \ll 1$, and we can similarly expand the exponential in the Fourier transform of $\bm{q}_1$:
\beq
\underset{\bm{q}_1}\int e^{i\bm{q}_1\cdot\bm{r}}f(\bm{q}_1)=\underset{\bm{q}_1}\int \Big(1+i\bm{q}_1\cdot\bm{r}+{\cal O}(\eta^2)\Big)f(\bm{q}_1)\,.
\eeq
This turns complicated Fourier transforms into integrals that resemble Feynman loop integrals, for which we can use established loop integration techniques. 
We illustrate this in the following examples. 

\subsection{Example 1: Non-Relativistic V Integral}
As a first example, we apply our technique on the known Fourier transform 
\beq\label{eq:ex1}
\begin{split}
I_1 &\equiv \underset{\bm{q}_1,\bm{q}_2,\bm{q}_3}\int e^{i\sum_{i=1}^3\bm{q}_i\cdot\bm{r}_i}\frac{(2\pi)^3\delta^{(3)}(\bm{q}_1+\bm{q}_2+\bm{q}_3)}{\bm{q}_1^2\bm{q}_2^2}
=\frac{1}{16\pi^2 r_{13} r_{23}}\,.
\end{split}
\eeq
This integral comes from a V-graph contribution (see Figure~\ref{fig:alldiags}) to the 6-point amplitude in the non-relativistic limit, where we drop the energy components of the momentum transfers, i.e., $q_i^0 \sim {\bm{p}_i \cdot \bm{q}_i \over m_i} \to 0$. 
Expanding in the hierarchical limit, \eqref{eq:ex1} becomes
\beq\label{eq:exp1V}
I_1=\frac{1}{16\pi^2}\frac{1}{|\bm{r}-\bm{\rho}|\rho}=\frac{1}{16\pi^2\rho^2\sqrt{1-2\eta\, x+\eta^2}}=\frac{1}{16\pi^2\rho^2}\sum_{k=0}^\infty \eta^k P_k(x)\,.
\eeq
where $x\equiv\hat{\bm{r}}\cdot\hat{\bm{\rho}}$, $\hat{\bm{r}}\equiv\bm{r}/r$, $\hat{\bm{\rho}}\equiv\bm{\rho}/\rho$, and $P_k(x)$ is the $k$th Legendre polynomial.
We will recover this expansion using the method of regions. 

Starting with the original integral in \eqref{eq:ex1}, we first eliminate $\bm{q}_2$ using the delta function
\beq
I_1 = \underset{\bm{q}_1,\bm{q}_3}\int \frac{e^{i\bm{q}_1\cdot\bm{r}}e^{i\bm{q}_3\cdot\bm{\rho}}}{\bm{q}_1^2(\bm{q}_1+\bm{q}_3)^2}\,.
\eeq
We then proceed to split the integral into regions a and b, expand in the small parameter $\eta$, compute the integral order by order in each region, and then resum the answer. In region a, we have $|\bm{q}_3|\sim 1/\rho \ll  |\bm{q}_1|\sim 1/r$, so we expand in small $|\bm{q}_3| / |\bm{q}_1| \sim \eta$. We find
\beq\label{eq:rega1}
\begin{split}
I_{1}\bigg|_{\textrm{a}} = \underset{\bm{q}_1, \bm{q}_3}\int \frac{e^{i\bm{q}_1\cdot\bm{r}}e^{i\bm{q}_3\cdot\bm{\rho}}}{\bm{q}_1^4}\bigg[&1-\frac{2(\bm{q}_1\cdot\bm{q}_3)}{\bm{q}_1^2}-\frac{\bm{q}_3^2}{\bm{q}_1^2}+\frac{4(\bm{q}_1\cdot\bm{q}_3)^2}{\bm{q}_1^4}+{\cal O} \left( \eta^3 \right)\bigg]\,.
\end{split}
\eeq
The Fourier transform factorized into a product of two decoupled Fourier transforms over $\bm{q}_1$ and $\bm{q}_3$ which can be computed using
\beq\label{eq:nonrelFT}
\begin{split}
\underset{\bm{q}}\int \frac{e^{i\bm{q}\cdot\bm{y}}\bm{q}^{i_1}\cdots \bm{q}^{i_n}}{\big(\bm{q}^2\big)^{\alpha}}&=(-i)^n\frac{\partial^n}{\partial\bm{y}_{i_1}\cdots\partial\bm{y}_{i_n}}\underset{\bm{q}}\int \frac{e^{i\bm{q}\cdot\bm{y}}}{\big(\bm{q}^2\big)^{\alpha}}\,,\\
\underset{\bm{q}}\int \frac{e^{i\bm{q}\cdot\bm{y}}}{\big(\bm{q}^2\big)^{\alpha}}&=\frac{\Gamma\big(d/2-\alpha\big)}{4^\alpha\pi^{d/2}\Gamma(\alpha)}\big(y^2\big)^{\alpha-d/2}\,,\qquad y>0
\end{split}
\eeq
in dimensional regularization with $d=3-2\epsilon$. For this example, we see in~\eqref{eq:rega1} that the pole in $\bm{q}_3$ has vanished, which means that the Fourier transform over $\bm{q}_3$ does not give a long-range potential as can be seen from~\eqref{eq:nonrelFT}. The contribution from region a therefore vanishes for $\rho>0$.

We now compute the contribution in region b, where $|\bm{q}_1|\sim 1/\rho$. We no longer have a hierarchy between $|\bm{q}_1|$ and $|\bm{q}_3|$, but we can expand the exponential for $\bm{q}_1\cdot\bm{r}\sim \eta \ll 1$:
\beq\label{eq:regb1}
\begin{split}
I_{1}\bigg|_{\textrm{b}} =\underset{\bm{q}_1,\bm{q}_3}\int \frac{e^{i\bm{q}_3\cdot\bm{\rho}}}{\bm{q}_1^2(\bm{q}_1+\bm{q}_3)^2}\bigg[&1+i\bm{q}_1\cdot\bm{r}-\frac{(\bm{q}_1\cdot\bm{r})^2}{2}+{\cal O}\left(\eta^3\right)\bigg]\,.
\end{split}
\eeq
The Fourier transform over $\bm{q}_1$ in region b has become a series of loop integrals which we can compute using Eq. (7.9) of~\cite{Bern:2019crd}:
\beq
\begin{split}
\underset{\bm{q}_1}\int \frac{\bm{q}_1^{i_1}\cdots\bm{q}_1^{i_n}}{\big(\bm{q}_1^2\big)^{\alpha}\big((\bm{q}_1+\bm{q}_3)^2\big)^{\beta}}&=\frac{(-1)^n}{(4\pi)^{d/2}\big(\bm{q}_3^2\big)^{\alpha+\beta-d/2}}\sum^{\lfloor n/2\rfloor}_{m=0}A(\alpha,\beta;n,m)\bigg(\frac{\bm{q}_3^2}{2}\bigg)^m\{[\delta]^m[\bm{q}_3]^{n-2m}\}^{i_1 \cdots i_n}\\
A(\alpha,\beta;n,m)&\equiv\frac{\Gamma(\alpha+\beta-m-d/2)\Gamma(n-m-\alpha+d/2)\Gamma(m-\beta+d/2)}{\Gamma(\alpha)\Gamma(\beta)\Gamma(n-\alpha-\beta+d)}
\end{split}
\eeq
for spatial indices $i_1,...,i_n$, where the object in curly brackets is the symmetric tensor made up of $m$ powers of the spatial metric and $n-2m$ powers of $\bm{q}_3$, normalized such that each unique term has unit coefficient. For our current example, we can evaluate \eqref{eq:regb1} using the simplified formula (Eq. (A.12) of~\cite{Smirnov:2006ry})
\beq
\begin{split}
\underset{\bm{q}_1}\int \frac{(\bm{q}_1\cdot\bm{r})^n}{\big(\bm{q}_1^2\big)^{\alpha}\big((\bm{q}_1+\bm{q}_3)^2\big)^{\beta}}&=\frac{(-1)^{n}}{(4\pi)^{d/2}\big(\bm{q}_3^2\big)^{\alpha+\beta-d/2}}\sum^{\floor*{n/2}}_{m=0}A(\alpha,\beta;n,m)\\
&\quad\times \frac{n!}{m!(n-2m)!}
\bigg(\frac{\bm{q}_3^2 r^2}{4}\bigg)^m(\bm{q}_3\cdot\bm{r})^{n-2m}\,.
\end{split}
\eeq
After computing the loop integral over $\bm{q}_1$, the Fourier transform over $\bm{q}_3$ can be performed using~\eqref{eq:nonrelFT}.
The result of the first few orders is 
\beq
\begin{split}
    I_1\bigg|_{\textrm{b}}&=\frac{1}{16\pi^2\rho^2}\bigg[1+\eta\, x+\frac{\eta^2}{2}(3x^2-1)+{\cal O}\big(\eta^3\big)\bigg]\,.
\end{split}
\eeq 
We see that this matches the first few orders in the expansion~\eqref{eq:exp1V}, and it is straightforward to verify at higher orders.

\subsection{Example 2: Non-Relativistic Y Integral}
We next compute
\beq\label{eq:ex2}
\begin{split}
I_2\equiv\underset{\bm{q}_1,\bm{q}_3}\int \frac{e^{i\bm{q}_1\cdot\bm{r}}e^{i\bm{q}_3\cdot\bm{\rho}}}{\bm{q}_1^2(\bm{q}_1+\bm{q}_3)^2\bm{q}_3^2}\,.
\end{split}
\eeq
Taking two partial derivatives of \eqref{eq:ex2} with respect to $\bm{r}$ or $\bm{\rho}$ yields contributions to the 6-point amplitude from the Y-graph (see Figure~\ref{fig:alldiags}) in the non-relativistic limit.

In region a, we perform the same expansion as in~\eqref{eq:rega1}, 
\beq\label{eq:rega2}
\begin{split}
I_2\bigg|_{\textrm{a}}&=\underset{\bm{q}_1,\bm{q}_3}\int \frac{e^{i\bm{q}_1\cdot\bm{r}}e^{i\bm{q}_3\cdot\bm{\rho}}}{\bm{q}_1^2\bm{q}_3^2}\frac{1}{\bm{q}_1^2}\bigg[1-\frac{2(\bm{q}_1\cdot\bm{q}_3)}{\bm{q}_1^2}-\frac{\bm{q}_3^2}{\bm{q}_1^2}+\frac{4(\bm{q}_1\cdot\bm{q}_3)^2}{\bm{q}_1^4}+{\cal O}\big(\eta^3 \big)\bigg]\,.
\end{split}
\eeq
As before, terms with no pole in $\bm{q}_3$ vanish. The remaining terms give a non-zero result in region a, given by
\beq\label{eq:I2a}
\begin{split}
    I_2\bigg|_{\textrm{a}}&=-\frac{\eta}{32\pi^2}-\frac{\eta^2 x}{64\pi^2}-\frac{\eta^3}{192\pi^2}(3x^2-1)-\frac{\eta^4}{256\pi^2}(5x^3-3x)+{\cal O}\big(\eta^5\big)\\
    &=-\frac{\eta}{32\pi^2}\sum_{k=0}^{\infty}\frac{\eta^kP_k(x)}{k+1} \,.
\end{split}
\eeq
The resummed result in the final line is obtained by recognizing the pattern order by order in $\eta$.

In region b, we perform the same loop integrals as in the previous example, and the Fourier transform over $\bm{q}_3$ is the same as before but with an additional factor of $\bm{q}_3^2$ in the denominator. The result is
\beq\label{eq:I2b}
I_2\bigg|_{\textrm{b}}=\frac{1}{32\pi^2}\bigg[-\log(\pi \rho^2)-\frac{1}{\epsilon}-\gamma\bigg]+\frac{1}{32\pi^2}\sum^{\infty}_{k=1}\frac{\eta^kP_k(x)}{k}\,.
\eeq
The full result for $I_2$ is given by summing the contributions from region a and b. Note that after taking any two partial derivatives with respect to $\bm r$ or $\bm \rho$ to get a classical contribution to the potential, the divergent pieces in \eqref{eq:I2b} drop out as expected.  We checked our result with~\cite{Loebbert:2020aos} up to ${\cal O}(\eta^5)$. We also applied our method on the integrals
\beq
\underset{\bm{q}_1,\bm{q}_3}\int \frac{e^{i\bm{q}_1\cdot\bm{r}}e^{i\bm{q}_3\cdot\bm{\rho}}}{\big(\bm{q}_1^2\big)^a\big((\bm{q}_1+\bm{q}_3)^2\big)^b\big(\bm{q}_3^2\big)^c}\,,\qquad (a,b,c)\in \Big\{(1,1,2),(2,2,1),(1,1,3)\Big\}\,,
\eeq
and checked the result with~\cite{Loebbert:2020aos}.\footnote{The dependence on the regularization prescription drops out after taking derivatives to get a classical contribution.}

\subsection{Example 3: Relativistic Y Integral}
Our next example is the relativistic version of $I_2$, 
\beq\label{eq:ex3}
\begin{split}
I_3\equiv\underset{\bm{q}_1,\bm{q}_3}\int \frac{e^{i\bm{q}_1\cdot\bm{r}}e^{i\bm{q}_3\cdot\bm{\rho}}}{[(\bm{v}_1\cdot\bm{q}_1)^2-\bm{q}_1^2][(\bm{v}_1\cdot\bm{q}_1+\bm{v}_3\cdot\bm{q}_3)^2-(\bm{q}_1+\bm{q}_3)^2][(\bm{v}_3\cdot\bm{q}_3)^2-\bm{q}_3^2]}\,,
\end{split}
\eeq
where $\bm{v}_i\equiv\bm{p}_i/E_i$.

In region a, 
\beq
\begin{split}
I_3\bigg|_\textrm{a}&=
    \underset{\bm{q}_1,\bm{q}_3}\int \frac{e^{i\bm{q}_1\cdot\bm{r}}e^{i\bm{q}_3\cdot\bm{\rho}}}{[(\bm{v}_1\cdot\bm{q}_1)^2-\bm{q}_1^2]^2[(\bm{v}_3\cdot\bm{q}_3)^2-\bm{q}_3^2]}\Bigg[1-2\frac{(\bm{v}_1\cdot\bm{q}_1)(\bm{v}_3\cdot\bm{q}_3)-\bm{q}_1\cdot\bm{q}_3}{(\bm{v}_1\cdot\bm{q}_1)^2-\bm{q}_1^2}+{\cal O}\big(\eta^2\big)\Bigg]\\
&=\frac{\gamma_1\gamma_3}{32\pi^2}\sqrt{\frac{1+\gamma_1^2(\hat{\bm{r}}\cdot\bm{v}_1)^2}{1+\gamma_3^2(\hat{\bm{\rho}}\cdot\bm{v}_3)^2}}
\begin{aligned}[t]
\Bigg\{\eta&+\frac{\eta^2}{2(1+\gamma_3^2(\hat{\bm{\rho}}\cdot\bm{v}_3)^2)}\bigg[\gamma_1^2(\hat{\bm{r}}\cdot\bm{v}_1)(\hat{\bm{\rho}}\cdot\bm{v}_1)+\hat{\bm{r}}\cdot\hat{\bm{\rho}}\\
&+\gamma_3^2(\hat{\bm{\rho}}\cdot\bm{v}_3)\Big(\gamma_1^2(\bm{v}_1\cdot\bm{v}_3-1)(\hat{\bm{r}}\cdot\bm{v}_1)+\hat{\bm{r}}\cdot\bm{v}_3\Big)\bigg]+{\cal O}\big(\eta^3\big)\Bigg\}
\end{aligned}
\end{split}
\eeq
with $\gamma_i\equiv(1-\bm{v}_i^2)^{-1/2}$. We obtained the above result using
\beq\label{eq:reldotft}
\begin{split}
G_\alpha(\bm{r},\bm{v}_i)&\equiv\underset{\bm{q}}\int \frac{e^{i\bm{q}\cdot\bm{r}}}{(q^2)^\alpha}=\underset{\bm{q}}\int \frac{e^{i\bm{q}\cdot\bm{r}}}{[(\bm{v}_i \cdot \bm{q})^2-\bm{q}^2]^\alpha}=\frac{(-1)^\alpha\Gamma(d/2-\alpha)\gamma_i }{\Gamma(\alpha)4^\alpha\pi^{d/2}}\Big(\bm{r}^2+\gamma_i^2(\bm{v}_i \cdot\bm{r})^2\Big)^{\alpha-d/2}
\end{split}
\eeq
where we set $q^0=\bm{v}_i\cdot \bm{q}$. We emphasize that the region a result is exact in velocity. 

In region b, 
\beq
\begin{split}
I_3\bigg|_\textrm{b}&=\underset{\bm{q}_1,\bm{q}_3}\int \frac{e^{i\bm{q}_3\cdot\bm{\rho}} \Big(1+i\bm{q}_1\cdot\bm{r}+{\cal O}\big(\eta^2\big)\Big)}{[(\bm{v}_1\cdot\bm{q}_1)^2-\bm{q}_1^2][(\bm{v}_1\cdot\bm{q}_1+\bm{v}_3\cdot\bm{q}_3)^2-(\bm{q}_1+\bm{q}_3)^2][(\bm{v}_3\cdot\bm{q}_3)^2-\bm{q}_3^2]}\, .
\end{split}
\eeq
The integral over $\bm{q}_1$ is now a series of loop integrals. One may proceed by computing in a PN expansion, $|\bm{v}_1|\sim|\bm{v}_3|\ll1$.

\section{Three-Body Potential in Position Space}\label{sec:hampos}
We present the conservative three-body interaction potential in position space at ${\cal O}(G^2)$ at LO and NLO in the hierarchical limit, which are respectively ${\cal O}\big({1 \over r^2}\big)$ and ${\cal O}\big({1 \over r \rho}\big)$. We derive the result using the method of regions outlined in Section~\ref{sec:mor}.

We begin by noting key properties of the contributions from regions a and b. In region b the exponential $e^{i \bm{q}_1 \cdot \bm{r}}$ is expanded, and the vector $\bm{r}$ thus factors out of the integral. This implies that the integral cannot produce inverse powers of $r$, and therefore it follows by dimensional analysis and power counting that the contribution from region b takes the following form
\beq
(\textrm{region b}) \sim {1 \over \rho^2} \Big( 1+ \eta + \eta^2 + \cdots \Big) \,.
\eeq
On the other hand, region a can produce inverse powers of $r$, and has the general form
\beq
(\textrm{region a}) \sim {1 \over r^2} \Big( 1+ \eta + \eta^2 + \cdots \Big) \,.
\eeq
Therefore the LO contribution at ${\cal O}\big(\frac{1}{r^2}\big)$ and NLO contribution at ${\cal O}\big(\frac{1}{r\rho}\big)$ can only come from region a. 
This reduces the complexity of the integrals immensely, and all necessary integrals factorize into the product of two Fourier transforms that can be computed exactly in velocity.

Moreover, the graphs that contribute at LO and NLO are those that factorize into a 5-point amplitude times a 3-point amplitude when $q_3$ is soft.
For example, Figure~\ref{fig:Vs} shows a graph that contributes at ${\cal O}\big(\frac{1}{r\rho}\big)$ and a graph that contributes at ${\cal O}\big(\frac{1}{\rho^2}\big)$. We can show this by analyzing the integrands in region a:
\beq
\begin{split}
    I_{\textrm{Fig. \ref{fig:Vs} left}}&\sim \underset{\bm{q}_1,\bm{q}_3}\int { e^{i \bm{q}_1 \cdot \bm{r}} e^{i \bm{q}_3 \cdot \bm{\rho}} \over q_1^2q_3^2}\sim\frac{1}{r\rho}\\
    I_{\textrm{Fig. \ref{fig:Vs} right}}&\sim \underset{\bm{q}_1,\bm{q}_3}\int { e^{i \bm{q}_1 \cdot \bm{r}} e^{i \bm{q}_3 \cdot \bm{\rho}} \over q_1^2 (q_1+q_3)^2} =\underset{\bm{q}_1,\bm{q}_3} \int {e^{i \bm{q}_1 \cdot \bm{r}} e^{i \bm{q}_3 \cdot \bm{\rho}} \over q_1^4} \Big( 1 + \cdots \Big) = 0\,,
\end{split}
\eeq
where the latter integral vanishes following the same argument used for~\eqref{eq:rega1}, i.e., using the Fourier transforms in~\eqref{eq:nonrelFT}.

\begin{figure}[t]
\centering
    \begin{tikzpicture}
        \draw[-] (0,0)--(2,0);
        \draw[-] (3,0)--(5,0);
        \draw[vector] (1,0)--(2.5,-3);
        \draw[vector] (4,0)--(2.5,-3);
        \draw[-] (1.5,-3)--(2.5,-3);
        \draw[-] (2.5,-3)--(3.5,-3);
        \node at (-.2,0) {1};
        \node at (1.3,-3) {2};
        \node at (2.8,0) {3};
    \end{tikzpicture}
    \hspace{1.25cm}
    \begin{tikzpicture}
        \draw[-] (0,0)--(2,0);
        \draw[-] (3,0)--(5,0);
        \draw[vector] (1,0)--(2.5,-3);
        \draw[vector] (4,0)--(2.5,-3);
        \draw[-] (1.5,-3)--(2.5,-3);
        \draw[-] (2.5,-3)--(3.5,-3);
        \node at (-.2,0) {1};
        \node at (1.3,-3) {3};
        \node at (2.8,0) {2};
    \end{tikzpicture}
    \caption{Example Feynman diagrams that contribute at ${\cal O}\big(\frac{1}{r\rho}\big)$ (left) and ${\cal O}\big(\frac{1}{\rho^2}\big)$ (right).}\label{fig:Vs}
\end{figure}

\subsection{V and Y Graph Contributions}
The largest contributions to the three-body potential from the V and Y graph topologies enter at ${\cal O}\big(\frac{1}{r\rho}\big)$. We therefore expand the momentum-space potential contributions from V and Y in region a
\beq\label{eq:VY1}
\begin{split}
\bigg(V_{\textrm{Y}}^{(3)}\big(\{\bm{p},\bm{q}\}\big)+V_{\textrm{V}}^{(3)}\big(\{\bm{p},\bm{q}\}\big) \bigg)\bigg|_{\textrm{a}}&=\frac{64 \pi^2 G^2 m_1^2 m_2^2 m_3^2}{E_1 E_2 E_3q_1^2 q_3^2}\\
&\quad\times\Bigg(1-4\sigma_{12}\sigma_{13}\sigma_{23}-\frac{(1-2\sigma_{12}^2)(p_3\cdot q_1)^2}{m_3^2q_1^2}+\mathcal{O}\left(\eta\right)\Bigg)
\end{split}
\eeq
where we used $p_2\cdot q_1=-p_2\cdot q_2 +{\cal O}(q_3)={\cal O}(\lambda^2)+{\cal O}(q_3) $. The Fourier transform of~\eqref{eq:VY1} is computed using~\eqref{eq:reldotft}. The result is 

\beq\label{eq:VY1r}
\begin{split}
V_{\textrm{Y}}^{(3)}\big(\{\bm{p},\bm{r}\}\big)+V_{\textrm{V}}^{(3)}\big(\{\bm{p},\bm{r}\}\big) 
&=\frac{4G^2 m_1 m_2^2 m_3}{E_2}\frac{1}{r\rho}\frac{1}{\sqrt{1+(\hat{\bm{r}}\cdot\bm{u}_1)^2}\sqrt{1+(\hat{\bm{\rho}}\cdot\bm{u}_3)^2}}\\
&\quad\times
\begin{aligned}[t]
\Bigg[1&-4\sigma_{12}\sigma_{13}\sigma_{23}+
\frac{(1-2\sigma_{12}^2)}{2(1+(\hat{\bm{r}}\cdot\bm{u}_1)^2)}\bigg(\sigma_{13}^2-1\\
&+2\sigma_{13}(\hat{\bm{r}}\cdot\bm{u}_1)(\hat{\bm{r}}\cdot\bm{u}_3)-(\hat{\bm{r}}\cdot\bm{u}_1)^2-(\hat{\bm{r}}\cdot\bm{u}_3)^2\bigg)+ {\cal O}\big(\eta\big)\Bigg]
\end{aligned}
\end{split}
\eeq
where we defined $\bm{u}_i\equiv\bm{p}_i/m_i = \gamma_i \bm{v}_i $, and used $\gamma_i = E_i/m_i$ and $\sigma_{ij}=\gamma_i\gamma_j(1-\bm{v}_i\cdot\bm{v}_j)$.

\subsection{Subtraction Contributions}\label{sec:sub}
We now consider the Fourier transform of the subtraction contributions given by
\beq\label{eq:Vsub}
\begin{split}
    V_{\textrm{sub},ijk}^{(3)}\big(\{\bm{p},\bm{q}\}\big)&=-V^{(2)}_{ij}V^{(2)}_{kj}\Bigg[
    \begin{aligned}[t]&\frac{1}{2E_j}-\frac{2\omega_{kj}^2}{E_{jk}q_{kj}^2}+\frac{E_j(q_i\cdot q_k)(\omega_i+\omega_{ij})(\omega_k+\omega_{kj})}{2E_{ij}q_i^2E_{jk}q_k^2}\\
    &+\frac{E_j}{E_{jk}^2q_k^2}\Bigg(2\omega_i\omega_{kj}+q_i\cdot q_k\bigg(1-\frac{(\omega_k+\omega_{kj})^2}{q_k^2}\bigg)\Bigg)\Bigg]\end{aligned}\\
    &\quad+16\pi G m_j m_k \sigma_{jk}V^{(2)}_{ij}\Bigg[\frac{1}{E_jq_{kj}^2}-\frac{p_k\cdot q_i}{E_kq_k^2}\Bigg(\frac{\omega_i+\omega_{ij}}{E_{ij}q_i^2}-\frac{\omega_k+\omega_{kj}}{E_{jk}q_{kj}^2}\Bigg)\Bigg] \,.
\end{split}
\eeq
In the hierarchical limit, terms with $j=3$ do not contribute at LO or NLO by power counting in region a. For terms with $j \neq 3$, the Fourier transform involves factors of the mismatched graviton poles
\beq\label{eq:mispoles}
\frac{1}{q_k^2q_{kj}^{2\alpha}}=\frac{1}{(\omega_k^2-\bm{q}_k^2)(\omega_{kj}^2-\bm{q}_k^2)^\alpha}\,,\qquad \alpha\in\{1,2\}\,.
\eeq
For $k\in\{1,2\}$, the mismatch in poles is higher order in the hierarchical limit, as can be seen from the following identities:
\beq
\omega_{12} = \omega_1 + {p_2 \cdot q_3 \over E_{12}} + {\cal O}(\lambda^2)  \,, \qquad \omega_{21} = \omega_2  + {p_1 \cdot q_3 \over E_{12}} + {\cal O}(\lambda^2) \,.
\eeq 
Therefore only the mismatched poles for $k=3$ remain. 

In the PN expansion, the Fourier transform to position space can be easily performed, and we can compare with the previous results for three-body potentials in Refs.~\cite{Will:2013cza} and~\cite{Loebbert:2020aos}. Our potential is in a different gauge, but we can instead compare gauge invariant amplitudes and we find agreement.

In the following sections, we describe two approaches to further simplify the Fourier transform integrals for the full relativistic case. The first approach is to take the planetary limit $m_3\gg m_1\sim m_2$, and the second approach is going to particle 3's rest frame.\footnote{We thank Callum Jones for noting a method for exactly calculating Fourier transforms involving the mismatched graviton poles in Eq. \eqref{eq:mispoles}. We leave this for future work.}

\subsubsection{Planetary Limit}
The planetary limit (PL) is defined by taking the mass of the distant third body to be much larger than the other two,
\beq
m_3 \gg m_1 \sim m_2 \sim m \,.
\eeq
This not only simplifies the Fourier transform of \eqref{eq:Vsub} but also describes a configuration relevant for astrophysics, such as a binary orbiting a supermassive black hole.  
We consider the bodies to have generic velocities, and take $m_i \sim |\bm{p}_i|\sim E_i$. In this limit, the remaining mismatched graviton poles can be aligned using the identity
\beq
\omega_{3j}=\omega_3 - {p_j \cdot q_3 \over E_{3j} } + {\cal O}(\lambda^2) \,,
\eeq
and then expanding in large $m_3$. 

Upon summing over the permutations and then expanding in the hierarchical limit and planetary limit, the subtraction contributions are given by
\beq\label{eq:VsubqPL2}
\begin{split}
V_{\textrm{sub}}^{(3)}\big(\{\bm{p},\bm{q}\}\big)\Big|_{\textrm{a, PL}} &= \sum_{(i,j,k)\in S_3} V_{\textrm{sub},ijk}^{(3)}\big(\{\bm{p},\bm{q}\}\big)\Big|_{\textrm{a, PL}} \\
&=\frac{16\pi^2G^2m_1^2m_2^2m_3^2}{E_1E_2E_3q_1^2q_3^2} \Bigg[c_1^{ij}\frac{\bm{q}_1^i \bm{q}_3^j}{q_3^2}+c_2+c_3^{ij}\frac{\bm{q}_1^i\bm{q}_1^j}{q_1^2}+c_4^{ij}\frac{\bm{q}_3^i\bm{q}_3^j}{q_3^2}
\\
&\qquad\quad+c_5^{ijkl}\frac{\bm{q}_1^i\bm{q}_1^j\bm{q}_3^k\bm{q}_3^l}{q_1^2q_3^2} +{\cal O}\bigg(\frac{m^2}{m_3^2}\bigg)+{\cal O}\big(\eta^2 \big)\Bigg]\,.
\end{split}
\eeq
In the second and third lines $i,j,k,l$ are spatial indices and the coefficients $c_n$ are given by
\beq
\begin{split}
    c_1^{ij}&=\frac{8(1-2\sigma_{12}^2)(m_1\sigma_{13}-m_2\sigma_{23})}{\gamma_3m_3}
    \bigg(\frac{\bm{u}_1^i\bm{u}_3^j}{\gamma_1}-\frac{\bm{u}_3^i\bm{u}_3^j}{\gamma_3}\bigg)\,,\\
    c_2&=\frac{4(\gamma_2\sigma_{12}(2\sigma_{13}^2-1)+\gamma_3\sigma_{13}(2\sigma_{12}^2-1))}{\gamma_1}- \frac{(2\sigma_{12}^2-1)(2\sigma_{13}^2-1)}{\gamma_1^2}+\\
    &\quad +\frac{4(\gamma_1\sigma_{12}(2\sigma_{23}^2-1)+\gamma_3 \sigma_{23}(2\sigma_{12}^2-1))}{\gamma_2} -\frac{(2\sigma_{12}^2-1)(2\sigma_{23}^2-1)}{\gamma_2^2}\,,\\
c_3^{ij}&=\frac{2(2\sigma_{12}^2-1)}{\gamma_1^2 E_{12}}\bigg[ 4\gamma_1 (m_1 \sigma_{13}+m_2 \sigma_{23})\bm{u}_3^i\bm{u}_1^j\\
&\quad+ \bigg( {m_1 \over \gamma_1} (2\sigma_{13}^2-1) + {m_2 \over \gamma_2} (2\sigma_{23}^2-1) -4\gamma_3 (m_1 \sigma_{13}+m_2 \sigma_{23})\bigg)\bm{u}_1^i\bm{u}_1^j\bigg]\,,\nonumber
\\
        c_4^{ij}&=\frac{8}{\gamma_3^2 m_3}\Bigg\{
        (2\sigma_{12}^2-1)(m_1\sigma_{13}+m_2\sigma_{23})\bm{u}_3^i\bm{u}_3^j\\
        &\quad+\bigg[\Big((2\sigma_{23}^2-1)\gamma_1m_2\sigma_{12}-(2\sigma_{13}^2-1)\gamma_2m_1\sigma_{12}-(2\sigma_{12}^2-1)\gamma_3m_1\sigma_{13}\Big)\\
        &\quad+\frac{m_1(2\sigma_{12}^2-1)(2\sigma_{13}^2-1)}{4\gamma_1}\bigg]\frac{\bm{u}_3^i\bm{u}_1^j}{\gamma_1}\\
        &\quad+\bigg[\Big((2\sigma_{13}^2-1)\gamma_2m_1\sigma_{12}-(2\sigma_{23}^2-1)\gamma_1m_2\sigma_{12}-(2\sigma_{12}^2-1)\gamma_3m_2\sigma_{23}\Big)\\
        &\quad+\frac{m_2(2\sigma_{12}^2-1)(2\sigma_{23}^2-1)}{4\gamma_2}\bigg]\frac{\bm{u}_3^i\bm{u}_2^j}{\gamma_2}
        \Bigg\}\,,
        \\
        c_5^{ijkl}&=\frac{4(2\sigma_{12}^2-1)}{\gamma_1^3\gamma_2\gamma_3^2m_3E_{12}}
        \Bigg[4\gamma_1^2\gamma_2^2m_2(m_1\sigma_{13}-m_2\sigma_{23})\bigg(\frac{\bm{u}_1^i\bm{u}_1^j\bm{u}_3^k\bm{u}_3^l}{\gamma_1}-\frac{\bm{u}_1^i\bm{u}_3^j\bm{u}_3^k\bm{u}_3^l}{\gamma_3}\bigg)\\
        &\quad+\gamma_2m_1^2(1 - 2 \sigma_{13}^2)\bm{u}_1^i\bm{u}_1^j\bm{u}_1^k\bm{u}_3^l+\gamma_1m_2^2(1 - 2 \sigma_{23}^2)\bm{u}_1^i\bm{u}_1^j\bm{u}_2^k\bm{u}_3^l\\
        &\quad+4\gamma_1\gamma_2\gamma_3\Big(m_1^2\sigma_{13}\bm{u}_1^i\bm{u}_1^j\bm{u}_1^k\bm{u}_3^l+m_2^2\sigma_{23}\bm{u}_1^i\bm{u}_1^j\bm{u}_2^k\bm{u}_3^l\Big)\\
        &\quad-4\gamma_1^2\gamma_2\Big(m_1^2\sigma_{13}\bm{u}_1^i\bm{u}_3^j\bm{u}_1^k\bm{u}_3^l+m_2^2\sigma_{23}\bm{u}_1^i\bm{u}_3^j\bm{u}_2^k\bm{u}_3^l\Big)\\
        &\quad-\gamma_1^2 \gamma_2\Big( m_1^2 (1-2 \sigma_{13}^2)+ m_2^2 (1-2 \sigma_{23}^2)+4m_1\gamma_3\sigma_{13}E_{12}\Big)\delta^{ik}\bm{u}_1^j\bm{u}_3^l\\
        &\quad+4m_1\gamma_1^3\gamma_2\sigma_{13}E_{12}\delta^{ik}\bm{u}_3^j\bm{u}_3^l\Bigg]\,.
\end{split}
\eeq
The contribution given by coefficient $c_1$ is LO in the hierarchical limit, while the contributions given by coefficients $c_2$, $c_3$, $c_4$, and $c_5$ are NLO. We have kept terms through ${\cal O}\left( {m/m_3} \right)$, i.e., NLO in the planetary limit. 

The Fourier transform of \eqref{eq:VsubqPL2} can be done using \eqref{eq:reldotft}. The result is 
\beq\label{eq:VsubrPL}
\begin{split}
V_{\textrm{sub}}^{(3)}\big(\{\bm{p},\bm{r}\}\big)\Big|_{\textrm{PL}}&=\frac{G^2m_1m_2^2m_3}{E_2}\frac{1}{\sqrt{1+(\hat{\bm{r}}\cdot\bm{u}_1)^2}\sqrt{1+(\hat{\bm{\rho}}\cdot\bm{u}_3)^2}r^2}\\
&\quad\times\Bigg[c_1^{ij}\frac{(\hat{\bm{r}}^i+(\bm{u}_1\cdot\hat{\bm{r}})\bm{u}_1^i)(\hat{\bm{\rho}}^j+(\bm{u}_3\cdot\hat{\bm{\rho}})\bm{u}_3^j)}{2(1+(\bm{u}_1\cdot\hat{\bm{r}})^2)}\\
&\quad+\eta\bigg(c_2+c_3^{ij}f^{ij}(\bm{r},\bm{u}_1)+c_4^{ij}f^{ij}(\bm{\rho},\bm{u}_3)+c_5^{ijkl}f^{ij}(\bm{r},\bm{u}_1)f^{kl}(\bm{\rho},\bm{u}_3)\bigg)\\
&\quad+{\cal O}\bigg(\frac{m^2}{m_3^2}\bigg) +{\cal O}\big(\eta^2\big)\Bigg]
\,,\\
f^{ij}(\bm{x},\bm{u})&\equiv\frac{\hat{\bm{x}}^i\hat{\bm{x}}^j-\bm{u}^i\bm{u}^j+(\hat{\bm{x}}\cdot\bm{u})(\hat{\bm{x}}^i\bm{u}^j+\hat{\bm{x}}^j\bm{u}^i)-(1+(\bm{u}\cdot\hat{\bm{x}})^2)\delta^{ij}}{2(1+(\bm{u}\cdot\hat{\bm{x}})^2)}\,.
\end{split}
\eeq
The full result in position space is given by the sum of \eqref{eq:VY1r} and \eqref{eq:VsubrPL}.

\subsubsection{Rest Frame of Particle 3}
Until now we have kept our results general for any choice of frame in position space: we have some generic set of 4-momenta $\{p_i^\mu=(E_i,\bm{p}_i)\}$ and use the on-shell condition to fix the energy component of the graviton momenta $q_i^0=\frac{\bm{p}_i\cdot\bm{q}_i}{E_i}=\omega_i$. Instead of taking $m_1\sim m_2\ll m_3$ to simplify the subtraction contribution integrals, we can instead go to the rest frame of particle 3, defined by
\beq
p_3^\mu=(m_3,0,0,0)
\eeq
and therefore
\beq
\omega_3=0\,,\qquad\omega_{3j}=\frac{\bm{p}_j\cdot\bm{q}_3}{m_3+E_j}\,,\qquad j\in \{1,2\}\,.
\eeq
The Fourier transform of Eq. \eqref{eq:mispoles} becomes
\beq
\begin{split}
\underset{\bm{q}_3}\int \frac{e^{i\bm{q}_3\cdot\bm{\rho}}}{(\omega_3^2-\bm{q}_3^2)(\omega_{3j}^2-\bm{q}_3^2)^\alpha}\bigg|_{\bm{p}_3=0}=-\underset{\bm{q}_3}\int \frac{e^{i\bm{q}_3\cdot\bm{\rho}}}{\bm{q}_3^2\Big[\Big(\frac{\bm{p}_j\cdot\bm{q}_3}{m_3+E_j}\Big)^2-\bm{q}_3^2\Big]^\alpha}\,.
\end{split}
\eeq
We compute this integral for $\alpha=1$ and $\alpha=2$:
\beq
\begin{split}
    J_\alpha(\bm{r},\bm{w})&\equiv \underset{\bm{q}}\int \frac{e^{i\bm{q}\cdot\bm{r}}}{\bm{q}^2\Big((\bm{w}\cdot\bm{q})^2-\bm{q}^2\Big)^\alpha}\\
    &=\frac{(-1)^{\alpha}\Gamma(1/2-\alpha)}{4^{\alpha+1}\pi^{3/2}\Gamma(\alpha)}\int_0^1dx\frac{x^{\alpha-1}\Big(\bm{r}^2+\frac{x(\bm{w}\cdot\bm{r})^2}{1-x\bm{w}^2}\Big)^{\alpha-1/2}}{\sqrt{1-x\bm{w}^2}}\,,
\end{split}
\eeq
\beq\label{eq:Js}
\begin{split}
\Rightarrow J_1(\bm{r},\bm{w})&=- \frac{r}{4\pi\bm{w}^2}\Bigg(\sqrt{1-x(1-c)}-\sqrt{c}\tanh^{-1}\bigg(\frac{\sqrt{1-x(1-c)}}{\sqrt{c}}\bigg)\Bigg)\Bigg|_{x=0}^{x=\bm{w}^2}\,,\\
J_2(\bm{r},\bm{w})&=\frac{r^3}{144\pi \bm{w}^4}
\begin{aligned}[t]\Bigg(&\frac{\sqrt{1-x(1-c)}}{1-x}\Big(2x^2(1-c)+2x(1-5c)-4+15c\Big)\\
&+3(3-5c)\sqrt{c}\tanh^{-1}\bigg(\frac{\sqrt{1-x(1-c)}}{\sqrt{c}}\bigg)\Bigg)\Bigg|_{x=0}^{x=\bm{w}^2}
\end{aligned}
    \end{split}
\eeq
with $c\equiv(\hat{\bm{r}}\cdot\hat{\bm{w}})^2$ and $\bm{w}^2<1$.

In momentum space, the subtraction contribution to the potential with $\bm{p}_3=0$ in region a is
\beq\label{eq:VsubqRF}
\begin{split}
V_{\textrm{sub}}^{(3)}
\left(\{\bm{p},\bm{q}\}\right)
\Big|_{\textrm{a, }\bm{p}_3=0} &= \sum_{(i,j,k)\in S_3} V_{\textrm{sub},ijk}^{(3)}\left(\{\bm{p},\bm{q}\}\right)\Big|_{\textrm{a, }\bm{p}_3=0} \\
&=\frac{16\pi^2G^2m_1^2m_2^2m_3}{E_1E_2} \\
&\quad\times\left[b_{11}^{ij}f_{11}^{ij}+b_{12}^{ij}f_{12}^{ij}+b_{13}^{ij}f_{13}^{ij}+b_{14}^{ij}f_{14}^{ij}+b_{21}^{ij}f_{21}^{ij}+b_{22}^{ij}f_{22}^{ij}+b_{23}^{ij}f_{23}^{ij}\right.\\
&\quad\left.+b_{24}^{ij}f_{24}^{ij}+b_{25}^{ijkl}f_{25}^{ijkl}+b_{26}^{ijkl}f_{26}^{ijkl}+b_{27}f_{27}+b_{28}f_{28}+b_{29}^{ij}f_{29}^{ij}+b_{210}^{ij}f_{210}^{ij}\right.\\
&\quad\left.+b_{211}^{ijkl}f_{211}^{ijkl}+b_{212}^{ijkl}f_{212}^{ijkl}+b_{213}^{ij}f_{213}^{ij}
\right]+\mathcal{O}\left(\frac{\eta^2}{r^2}\right)
\end{split}
\eeq
and the position-space result from the subtraction terms at LO and NLO in the hierarchical limit is 
\beq\label{eq:VsubrRF}
\begin{split}
V_{\textrm{sub}}^{(3)}
\left(\{\bm{p},\bm{r}\}\right)
\Big|_{\bm{p}_3=0} &=\frac{16\pi^2G^2m_1^2m_2^2m_3}{E_1E_2} \\
&\quad\times\left[b_{11}^{ij}g_{11}^{ij}+b_{12}^{ij}g_{12}^{ij}+b_{13}^{ij}g_{13}^{ij}+b_{14}^{ij}g_{14}^{ij}+b_{21}^{ij}g_{21}^{ij}+b_{22}^{ij}g_{22}^{ij}+b_{23}^{ij}g_{23}^{ij}\right.\\
&\quad\left.+b_{24}^{ij}g_{24}^{ij}+b_{25}^{ijkl}g_{25}^{ijkl}+b_{26}^{ijkl}g_{26}^{ijkl}+b_{27}g_{27}+b_{28}g_{28}+b_{29}^{ij}g_{29}^{ij}+b_{210}^{ij}g_{210}^{ij}\right.\\
&\quad\left.+b_{211}^{ijkl}g_{211}^{ijkl}+b_{212}^{ijkl}g_{212}^{ijkl}+b_{213}^{ij}g_{213}^{ij}
\right]+\mathcal{O}\left(\frac{\eta^2}{r^2}\right)\,.
\end{split}
\eeq
The terms with coefficients $b_{1\textrm{x}}$ are LO in the hierarchical limit, and the terms with coefficients $b_{2\textrm{x}}$ are NLO in the hierarchical limit.
The coefficients $b$ and functions $f$, $g$ are listed in appendix~\ref{sec:appRF}. As in Eqs.~\eqref{eq:VsubqPL2}~and~\eqref{eq:VsubrPL}, the superscripts in the second equality of Eq.~\eqref{eq:VsubqRF}~and in~\eqref{eq:VsubrRF} denote spatial indices. 
The full potential is given by the sum of \eqref{eq:VsubrRF} and \eqref{eq:VY1r} with $\bm{p}_3=0$.

\section{Sequence of EFTs}\label{sec:EFTs}

\begin{figure}[t]\centering
\begin{tikzpicture}[roundnode/.style={circle, draw=black, very thick, minimum size=3mm},cross/.style={thick,fill=white,path picture={ 
  \draw[black]
(path picture bounding box.south east) -- (path picture bounding box.north west) (path picture bounding box.south west) -- (path picture bounding box.north east);
}}]
\draw [-,thick] (0,10) -- (15,10);
\draw[-,thick] (15/4,9)--(15/4,11);
\draw[-,thick] (15/4*3,9)--(15/4*3,11);
\node at (15/4,11.3) {$r$};
\node at (15/4*3,11.3) {$\rho$};
\draw[dashed,thick] (15/4,0) -- (15/4,9);
\draw[dashed,thick] (15/4*3,0) -- (15/4*3,9);
\node at (15/8,10.5) {Full Theory};
\draw[-,thick] (15/8-1,9)--(15/8+1,9);
\draw[-,thick] (15/8-1,9-1)--(15/8+1,9-1);
\draw[vector,thick,blue] (15/8+0.05,9-.5)--(15/8+1,9-.5);
\draw[vector,thick,orange] (15/8,9)--(15/8,9-1);
\node at (15/8-1.2,9) {$1$};
\node at (15/8-1.2,9-1) {$2$};
\node at (15/8+1.3,9-.5) {$q_3$};
\draw[-,thick] (15/8-1,9-2)--(15/8+1,9-2);
\draw[-,thick] (15/8-1,9-1-2)--(15/8+1,9-1-2);
\draw[vector,thick,blue] (15/8+0.4,9-1-2)--(15/8+1,9-.5-2);
\draw[vector,thick,orange] (15/8,9-2)--(15/8,9-1-2);
\node at (15/8-1.2,9-2) {$1$};
\node at (15/8-1.2,9-1-2) {$2$};
\node at (15/8+1.3,9-.5-2) {$q_3$};
\node at (15/2,10.5) {Inner-Binary EFT};
\node at (15/8*7,10.5) {Three-Body EFT};
\draw[-,thick] (15/8*3-1,9)--(15/8*3+1,9-1);
\draw[-,thick] (15/8*3-1,9-1)--(15/8*3+1,9);
\filldraw[fill=black, thick](15/8*3,9-.5) circle (0.1);
\draw[vector,thick,blue] (15/8*3+.1,9-.5)--(15/8*3+1,9-.5);
\node at (15/8*3-1.2,9) {$1$};
\node at (15/8*3-1.2,9-1) {$2$};
\node at (15/8*3+1.3,9-.5) {$q_3$};
\draw[-,thick] (15/8*3-1,9-2)--(15/8*3+1,9-1-2);
\draw[-,thick] (15/8*3-1,9-1-2)--(15/8*3+1,9-2);
\filldraw[fill=black, thick](15/8*3,9-.5-2) circle (0.1);
\draw[vector,thick,blue] (15/8*3+.4,9-.7-2)--(15/8*3+1,9-.5-2);
\node at (15/8*3-1.2,9-2) {$1$};
\node at (15/8*3-1.2,9-1-2) {$2$};
\node at (15/8*3+1.3,9-.5-2) {$q_3$};
\filldraw[fill=gray!20,rounded corners,draw=gray!30](15/8*5-1.7,9-3.6) rectangle (15/8*5+1.7,9.8);
\node at (15/8*5,9.5) {region a};
\draw[-,thick] (15/8*5-1-.25,9)--(15/8*5+1-.25,9-1);
\draw[-,thick] (15/8*5-1-.25,9-1)--(15/8*5+1-.25,9);
\filldraw[fill=black, thick](15/8*5-.25,9-.5) circle (0.1);
\draw[vector,thick,blue] (15/8*5+.1-.25,9-.5)--(15/8*5+1.5-.125,9-.5);
\draw[-,thick] (15/8*5+1.5-.25,9) -- (15/8*5+1.5,9-1);
\node at (15/8*5-1.2-.25,9) {$1$};
\node at (15/8*5-1.2-.25,9-1) {$2$};
\node at (15/8*5+1.5,9-1.3) {$3$};
\draw[-,thick] (15/8*5-1-.25,9-2)--(15/8*5+1-.25,9-1-2);
\draw[-,thick] (15/8*5-1-.25,9-1-2)--(15/8*5+1-.25,9-2);
\filldraw[fill=black, thick](15/8*5-.25,9-.5-2) circle (0.1);
\draw[vector,thick,blue] (15/8*5+.4-.25,9-.7-2)--(15/8*5+1.5-.125,9-.5-2);
\draw[-,thick] (15/8*5+1.5-.25,9-2) -- (15/8*5+1.5,9-1-2);
\node at (15/8*5-1.2-.25,9-2) {$1$};
\node at (15/8*5-1.2-.25,9-1-2) {$2$};
\node at (15/8*5+1.5,9-1.3-2) {$3$};
\filldraw[fill=gray!20,rounded corners,draw=gray!30](15/8*5-1.7,9-3.6-5) rectangle (15/8*5+1.7,9.8-5);
\node at (15/8*5,9.5-5) {region b};
\draw[-,thick] (15/8*5-1-.25,9-5)--(15/8*5-.25+1,9-5);
\draw[-,thick] (15/8*5-1-.25,9-1-5)--(15/8*5-.25+1,9-1-5);
\draw[vector,thick,blue] (15/8*5-.25+0.05,9-.5-5)--(15/8*5+1.5-.125,9-.5-5);
\draw[vector,thick,blue] (15/8*5-.25,9-5)--(15/8*5-.25,9-1-5);
\draw[-,thick] (15/8*5+1.5-.25,9-5) -- (15/8*5+1.5,9-1-5);
\draw[-,thick] (15/8*5-1-.25,9-5-2)--(15/8*5-.25+1,9-5-2);
\draw[-,thick] (15/8*5-1-.25,9-1-5-2)--(15/8*5-.25+1,9-1-5-2);
\draw[vector,thick,blue] (15/8*5-.25+0.4,9-1-5-2)--(15/8*5+1.5-.125,9-.5-5-2);
\draw[vector,thick,blue] (15/8*5-.25,9-5-2)--(15/8*5-.25,9-1-5-2);
\draw[-,thick] (15/8*5+1.5-.25,9-5-2) -- (15/8*5+1.5,9-1-5-2);
\draw[-,thick] (15/8*7-1,9-1)--(15/8*7+1,9);
\draw[-,thick] (15/8*7-1,9-.5)--(15/8*7+1,9-.5);
\draw[-,thick] (15/8*7-1,9)--(15/8*7+1,9-1);
\filldraw[fill=black, thick](15/8*7,9-.5) circle (0.1);
\draw[-,thick] (15/8*7-1,9-2)--(15/8*7+1,9-1-2);
\draw[-,thick] (15/8*7-1,9-2-.5)--(15/8*7+1,9-2);
\draw[-,thick] (15/8*7-1,9-2-1)--(15/8*7+1,9-2-.5);
\filldraw[fill=black, thick] (15/8*7-.33,9-2-.33) circle (0.1);
\filldraw[fill=black, thick] (15/8*7+.34,9-2-.66) circle (0.1);
\end{tikzpicture}
\caption{Sample diagrams illustrating the matching to a sequence of EFTs in the hierarchical limit. At scale $r$, potential gravitons with momentum $\sim\frac{1}{r}$ (orange wavy line) are integrated out. The full theory includes 5-point graphs where the exchanged graviton has momentum $\sim\frac{1}{r}$ (orange wavy line) and the emitted graviton has soft momentum $\sim\frac{1}{\rho}$ (blue wavy line). These are matched to contact terms (black circles) in the Inner-Binary EFT. At scale $\rho$, potential gravitons with momentum $\sim\frac{1}{\rho}$ (blue wavy line) are integrated out. In the Inner-Binary EFT, this includes graphs where particle 3 is attached to the 5-point graphs, as well as 6-point graphs where all exchanged gravitons have momentum $\frac{1}{\rho}$. These are  matched to contact diagrams in the Three-Body EFT. We also match 4-point diagrams at both scale $r$ and $\rho$ to determine two-body potentials, but these are not explicitly shown here.}
\label{fig:efts}
\end{figure}

Our analysis using the method of regions can also be implemented by matching to a sequence of EFTs; see Figure~\ref{fig:efts}. The EFT at scale $r$ captures contributions from the physics of the inner binary composed of particles 1 and 2, and we refer to this as the Inner-Binary EFT. The EFT at scale $\rho$ captures long-distance interactions of the inner binary with particle 3, as well as long-distance contributions among all three particles; we refer to this as the Three-Body EFT. Matching is performed using generalized isotropic gauge, and, similar to the analysis presented above, there are iteration terms that cancel in the matching and corresponding subtraction contributions. The basic mechanics of this matching is described below.

At scale $r$, we determine the Inner-Binary EFT by integrating out potential gravitons with momentum of order $\frac{1}{r}$ that mediate interactions between particles 1 and 2. This involves matching 4-point diagrams for the scattering of particles 1 and 2 onto an effective potential, and matching 5-point diagrams for the scattering of particles 1 and 2 with the emission of a soft graviton onto a 5-point contact term describing an effective stress-tensor for the inner binary. The soft graviton has momentum $q_3\sim\frac{1}{\rho}\ll\frac{1}{r}$.

At scale $\rho$, we determine the Three-Body EFT by integrating out potential gravitons with momentum of order $\frac{1}{\rho}$. This involves matching both 4-point and 6-point amplitudes between the two EFTs. The contributions from the effective stress-tensor of the inner binary correspond to region a, where the Fourier transform factorizes. On the other hand, contributions where all gravitons have momentum of order $\frac{1}{\rho}$ do not factorize in this way and correspond to region b.

In this EFT analysis, higher-order corrections of order $\frac{G m_{1,2}}{r}$ or $\frac{G m_3}{\rho}$ can be separately targeted as loops in either the Inner-Binary EFT or the Three-Body EFT. Moreover, renormalization group effects that connect these two EFTs would capture $\log\left(r/\rho\right)$ terms.

Our method of regions directly connects to the analysis presented in \cite{Georgoudis:2023eke}. In that paper, the authors take the soft limit of the gravitational waveform $\omega\rightarrow 0$ (our ``soft'' $q_3$) and consider the Fourier transform of $q_\perp$ (our $q_1$) to impact parameter space $b$ (our $r$). Their two regions are $\omega\ll q_\perp\sim \frac{1}{b}$ (our region a) and $\omega\sim q_\perp\ll \frac{1}{b}$ (our region b). The expansion in each region is the same as ours, but their analysis applies this for a different observable and thus derives different integrals.

\section{Discussion}\label{sec:discussion}
The main result of this paper is the position-space three-body ${\cal O}(G^2)$ conservative potential at LO and NLO in the hierarchical limit, calculated in the rest frame of the distant particle or in the limit where the distant particle is much heavier than the other two particles (in a general frame). The hierarchical limit not only allows us to systematically derive analytic results at all orders in velocity but also describe a three-body configuration relevant for astrophysical studies. These developments again show how amplitudes- and EFT-based methods can be leveraged for classical GR calculations. 

Our new analytic results (Eqs.~\eqref{eq:VY1r},~\eqref{eq:VsubrPL},~and~\eqref{eq:VsubrRF}) may be used to model both bound and unbound three-body systems in GR. An exciting prospect would be to use our potential, together with the two-body ${\cal O}(G^2)$ potential (calculated in a general frame in~\cite{Lee:2023zuu}) to derive equations of motion in unbound hierarchical triple systems, and model the ejection of a hypervelocity star or a tidal distruption event within the context of hierarchical triples.
Another natural application is to study the effect of three-body interactions on binary dynamics and GWs, and the implications for the quantity and characteristics of eccentric binaries detectable with LISA~\cite{LISA:2022yao,LISAConsortiumWaveformWorkingGroup:2023arg,Ficarra:2023zjc}. 

Results in momentum space are certainly important, but in order to be useful to the wider GR and astrophysics community they should be transformed into position space and expressed in an appropriate coordinate system. In this work we initiate the application of the method of regions for evaluating the Fourier transforms, providing a prescription for applying position-space constraints directly in momentum space as well as insights into the physics of three-body systems. In particular, the contributions from different regions can be systematically studied using an EFT setup that separately captures the physics at energy scales $1/r$ and $1/\rho$. It would be interesting to study this EFT in detail, e.g., how properties of the inner binary are captured, how interactions between the inner binary and the third distant body are described as an effective two-body system, and how properties of the five-point amplitude, such as its soft limit, are encoded in the EFT. 

There are many other avenues for future work. It would be interesting to explore how the hierarchical limit and method of regions can be applied to broader classes of integrals such as ``triple-K integrals''~\cite{Jones:2022aji,Bzowski:2020lip}, and for deriving results at higher orders in $G$. Moreover, a more careful study of gauge and coordinate ambiguities would be highly useful. It would also be interesting to include the effects of spin, radiation, and finite sizes, and to study simplified three-body configurations where observable quantities such as scattering angles can be studied in detail. In particular, it would be interesting to consider the effect of the distant third body on the scattering angle of two bodies, or on the waveform produced by the inner binary. It would also be interesting to employ the KMOC formalism~\cite{Kosower:2018adc} to directly compute the momentum impulses of the three particles.

\acknowledgments
The authors would like to thank Callum Jones, Dimitrios Kosmopoulos,  Trevor Scheopner, and Smadar Naoz for useful discussions.
M.~P.~S.~is supported by the US Department of Energy Early Career program under award number DE-SC0024224, and the Sloan Foundation. 
A.~M.~W.~is supported by the NSF Graduate Research Fellowship under Grant No. DGE-2034835. We are also grateful to the Mani L. Bhaumik Institute for Theoretical Physics for support.

\newpage
\appendix
\section{Rest Frame of Particle 3 Functions}\label{sec:appRF}
The coefficients of the 3-body potential in the rest frame of the distant particle (Eqs.~\eqref{eq:VsubqRF}~and~\eqref{eq:VsubrRF}) are 
\begin{align}
    b_{11}^{ij}&=-\frac{2 \left(2 \sigma_{12}^2-1\right) \left(2 E_1^2-m_1^2\right) \left((E_1+m_3) \delta^{ij}-E_1\bm{v}_1^i\bm{v}_1^j\right)}{(E_1+m_3)^3}\nonumber\\
    b_{12}^{ij}&=-\frac{\left(2 \sigma_{12}^2-1\right) \left(\left(2 E_1^2-m_1^2\right) \delta^{ij}-4E_1^2 \bm{v}_1^i\bm{v}_1^j\right)}{(E_1+m_3)^2}\nonumber\\
    b_{13}^{ij}&=\frac{2 \left(2 \sigma_{12}^2-1\right) \left(2 E_2^2-m_2^2\right) \left((E_2+m_3) \delta^{ij}-E_2\bm{v}_1^i\bm{v}_2^j\right)}{(E_2+m_3)^3}\nonumber\\
    b_{14}^{ij}&=\frac{\left(2 \sigma_{12}^2-1\right) \left(\left(2 E_2^2-m_2^2\right) \delta^{ij}-4E_2^2 \bm{v}_1^i\bm{v}_2^j\right)}{(E_2+m_3)^2}\nonumber\\
    b_{21}^{ij}&=\frac{2 \left(2 \sigma_{12}^2-1\right) \left(2 E_1^2-m_1^2\right) \bm{v}_1^i\bm{v}_1^j}{E_1(E_1+E_2)}\nonumber\\
    b_{22}^{ij}&=\frac{2 \left(2 \sigma_{12}^2-1\right) \left(2 E_2^2-m_2^2\right) \bm{v}_1^i\bm{v}_1^j}{E_2 (E_1+E_2)}\nonumber\\
    b_{23}^{ij}&=\frac{4 E_1 E_2 m_1 \sigma_{12} \left(m_1^2-2 E_1^2\right)\bm{v}_1^i\bm{v}_2^j}{m_1^2m_2  (E_1+m_3)^2}\nonumber\\
    b_{24}^{ij}&=\frac{4 E_1 E_2 m_2 \sigma_{12} \left(m_2^2-2 E_2^2\right) \bm{v}_1^i\bm{v}_2^j}{m_1 m_2^2(E_2+m_3)^2}\nonumber\\
    b_{25}^{ijkl}&=-\frac{2 \left(2 \sigma_{12}^2-1\right) \left(4 E_1^3 \bm{v}_1^i\bm{v}_1^j\bm{v}_1^k\bm{v}_1^l+(E_1+E_2) \left(2 E_1^2-m_1^2\right) \delta^{ik}\delta^{jl}-4 E_1^2 (E_1+E_2) \delta^{ik}\bm{v}_1^j \bm{v}_1^l\right)}{(E_1+E_2) (E_1+m_3)^2}\nonumber\\
    b_{26}^{ijkl}&=\frac{8 E_2^3 \left(1-2 \sigma_{12}^2\right) \bm{v}_1^i\bm{v}_1^j\bm{v}_2^k\bm{v}_2^l}{(E_1+E_2) (E_2+m_3)^2}\nonumber\\
    b_{27}&=\frac{4 E_1 E_2 \sigma_{12} \left(2 E_1^2-m_1^2\right) (E_1+m_3)^2+m_1 m_2 \left(2 \sigma_{12}^2-1\right) \left(4 E_1^4+8 E_1^3 m_3+m_3^2 \left(2 E_1^2+m_1^2\right)\right)}{E_1^2 m_1 m_2 (E_1+m_3)^2}\nonumber\\
    b_{28}&=\frac{4 E_1 E_2 \sigma_{12} \left(2 E_2^2-m_2^2\right) (E_2+m_3)+m_1m_2\left(2 \sigma_{12}^2-1\right)\left(6 E_2^3+2 E_2^2 m_3-E_2 m_2^2+m_2^2 m_3\right)}{E_2^2 m_1 m_2  (E_2+m_3)}\nonumber\\
    b_{29}^{ij}&=\frac{2 m_3  \left(2 \sigma_{12}^2-1\right) \left(2 E_1^2-m_1^2\right)\delta^{ij}}{E_1 (E_1+m_3)^2}\nonumber\\
    b_{210}^{ij}&=\frac{2\left(2 \sigma_{12}^2-1\right) \left(2 E_2^2-m_2^2\right)\delta^{ij}}{E_2  (E_2+m_3)}\nonumber\\
    b_{211}^{ijkl}&=-\frac{4 \left(2 \sigma_{12}^2-1\right) \left(2 E_1^2-m_1^2\right) \left((E_1+E_2) \delta^{ik}-E_1 \bm{v}_1^i\bm{v}_1^k\right) \left((E_1+m_3) \delta^{jl}-E_1\bm{v}_1^j\bm{v}_1^l\right)}{(E_1+E_2) (E_1+m_3)^3}\nonumber\\
    b_{212}^{ijkl}&=\frac{4 E_2 \left(2 \sigma_{12}^2-1\right) \left(2 E_2^2-m_2^2\right) \bm{v}_1^i\bm{v}_2^k \left((E_2+m_3) \delta^{jl}-E_2 \bm{v}_1^j\bm{v}_2^k\right)}{ (E_1+E_2) (E_2+m_3)^3}\nonumber\\
    b_{213}^{ij}&=8 \left(2 \sigma_{12}^2-1\right) \bm{v}_1^i\bm{v}_1^j
\end{align}
The functions in momentum space are 
\begin{align}\label{eq:fs1}
    f_{11}^{ij}&=\frac{\bm{q}_1^i\bm{q}_3^j}{q_1^2\left(\left(\frac{E_1 \bm{v}_1\cdot\bm{q}_3}{E_1+m_3}\right)^2-\bm{q}_3^2\right)^2} & f_{12}^{ij}&=\frac{\bm{q}_1^i\bm{q}_3^j}{q_1^2\bm{q}_3^2\left(\left(\frac{E_1 \bm{v}_1\cdot\bm{q}_3}{E_1+m_3}\right)^2-\bm{q}_3^2\right)}\nonumber\\
    f_{13}^{ij}&=\frac{\bm{q}_1^i\bm{q}_3^j}{q_1^2\left(\left(\frac{E_2 \bm{v}_2\cdot\bm{q}_3}{E_2+m_3}\right)^2-\bm{q}_3^2\right)^2} & f_{14}^{ij}&=\frac{\bm{q}_1^i\bm{q}_3^j}{q_1^2\bm{q}_3^2\left(\left(\frac{E_2 \bm{v}_2\cdot\bm{q}_3}{E_2+m_3}\right)^2-\bm{q}_3^2\right)}\nonumber\\
    f_{21}^{ij}&=\frac{\bm{q}_1^i\bm{q}_1^j}{q_1^4\left(\left(\frac{E_1 \bm{v}_1\cdot\bm{q}_3}{E_1+m_3}\right)^2-\bm{q}_3^2\right)}
        &
        f_{22}^{ij}&=\frac{\bm{q}_1^i\bm{q}_1^j}{q_1^4\left(\left(\frac{E_2 \bm{v}_2\cdot\bm{q}_3}{E_2+m_3}\right)^2-\bm{q}_3^2\right)}
        \nonumber\\
        f_{23}^{ij}&=\frac{\bm{q}_3^i\bm{q}_3^j}{q_1^2\bm{q}_3^2\left(\left(\frac{E_1 \bm{v}_1\cdot\bm{q}_3}{E_1+m_3}\right)^2-\bm{q}_3^2\right)}
        &
        f_{24}^{ij}&=\frac{\bm{q}_3^i\bm{q}_3^j}{q_1^2\bm{q}_3^2\left(\left(\frac{E_2 \bm{v}_2\cdot\bm{q}_3}{E_2+m_3}\right)^2-\bm{q}_3^2\right)}
        \nonumber\\
        f_{25}^{ijkl}&=\frac{\bm{q}_1^i\bm{q}_1^j\bm{q}_3^k\bm{q}_3^l}{q_1^4\bm{q}_3^2\left(\left(\frac{E_1 \bm{v}_1\cdot\bm{q}_3}{E_1+m_3}\right)^2-\bm{q}_3^2\right)}
        &
        f_{26}^{ijkl}&=\frac{\bm{q}_1^i\bm{q}_1^j\bm{q}_3^k\bm{q}_3^l}{q_1^4\bm{q}_3^2\left(\left(\frac{E_2 \bm{v}_2\cdot\bm{q}_3}{E_2+m_3}\right)^2-\bm{q}_3^2\right)}
        \nonumber\\
        f_{27}&=\frac{1}{q_1^2\left(\left(\frac{E_1 \bm{v}_1\cdot\bm{q}_3}{E_1+m_3}\right)^2-\bm{q}_3^2\right)}
        &
        f_{28}&=\frac{1}{q_1^2\left(\left(\frac{E_2 \bm{v}_2\cdot\bm{q}_3}{E_2+m_3}\right)^2-\bm{q}_3^2\right)}
        \nonumber\\
        f_{29}^{ij}&=\frac{\bm{q}_3^i\bm{q}_3^j}{q_1^2\left(\left(\frac{E_1 \bm{v}_1\cdot\bm{q}_3}{E_1+m_3}\right)^2-\bm{q}_3^2\right)^2}
        &
        f_{210}^{ij}&=\frac{\bm{q}_3^i\bm{q}_3^j}{q_1^2\left(\left(\frac{E_2 \bm{v}_2\cdot\bm{q}_3}{E_2+m_3}\right)^2-\bm{q}_3^2\right)^2}
        \nonumber\\
        f_{211}^{ijkl}&=\frac{\bm{q}_1^i\bm{q}_1^j\bm{q}_3^k\bm{q}_3^l}{q_1^4\left(\left(\frac{E_1 \bm{v}_1\cdot\bm{q}_3}{E_1+m_3}\right)^2-\bm{q}_3^2\right)^2}
        &
        f_{212}^{ijkl}&=\frac{\bm{q}_1^i\bm{q}_1^j\bm{q}_3^k\bm{q}_3^l}{q_1^4\left(\left(\frac{E_2 \bm{v}_2\cdot\bm{q}_3}{E_2+m_3}\right)^2-\bm{q}_3^2\right)^2}
        \nonumber\\
        f_{213}^{ij}&=\frac{\bm{q}_1^i\bm{q}_1^j}{q_1^4\bm{q}_3^2}
\end{align}

We define the Fourier transforms to be
\beq
g_{\alpha}^{i_1\cdots i_n}\equiv\underset{\bm{q}_1,\bm{q}_3}\int e^{i\bm{q}_1\cdot\bm{r}}e^{i\bm{q}_3\cdot\bm{\rho}}f_\alpha^{i_1\cdots i_n}
\eeq
where the subscript $\alpha$ runs over the labels of the $f$ functions in Eq.~\eqref{eq:fs1} with the appropriate Lorentz spatial indices $i_1\cdots i_n$. 

Using Eqs.~\eqref{eq:reldotft}~and~\eqref{eq:Js}, the $g$'s are therefore
\beq
\begin{split}
    g_{11}^{ij}&= -\frac{\partial^2}{\partial\bm{r}_i\partial\bm{\rho}_j}G_1(\bm{r},\bm{v}_1)G_2\left(\bm{\rho},\frac{E_1 \bm{v}_1}{E_1+m_3}\right)\\
    g_{12}^{ij}&=- \frac{\partial^2}{\partial\bm{r}_i\partial\bm{\rho}_j}G_1(\bm{r},\bm{v}_1)J_1\left(\bm{\rho},\frac{E_1 \bm{v}_1}{E_1+m_3}\right)\\
    g_{13}^{ij}&= -\frac{\partial^2}{\partial\bm{r}_i\partial\bm{\rho}_j}G_1(\bm{r},\bm{v}_1)G_2\left(\bm{\rho},\frac{E_2 \bm{v}_2}{E_2+m_3}\right)\\
    g_{14}^{ij}&=- \frac{\partial^2}{\partial\bm{r}_i\partial\bm{\rho}_j}G_1(\bm{r},\bm{v}_1)J_1\left(\bm{\rho},\frac{E_2 \bm{v}_2}{E_2+m_3}\right)\\
    g_{21}^{ij}&=- \frac{\partial^2}{\partial\bm{r}_i\partial\bm{r}_j}G_2(\bm{r},\bm{v}_1)G_1\left(\bm{\rho},\frac{E_1 \bm{v}_1}{E_1+m_3}\right)\\
    g_{22}^{ij}&= -\frac{\partial^2}{\partial\bm{r}_i\partial\bm{r}_j}G_2(\bm{r},\bm{v}_1)G_1\left(\bm{\rho},\frac{E_2 \bm{v}_2}{E_2+m_3}\right)\\
    g_{23}^{ij}&= -\frac{\partial^2}{\partial\bm{\rho}_i\partial\bm{\rho}_j}G_1(\bm{r},\bm{v}_1)J_1\left(\bm{\rho},\frac{E_1 \bm{v}_1}{E_1+m_3}\right)\\
    g_{24}^{ij}&=- \frac{\partial^2}{\partial\bm{\rho}_i\partial\bm{\rho}_j}G_1(\bm{r},\bm{v}_1)J_1\left(\bm{\rho},\frac{E_2 \bm{v}_2}{E_2+m_3}\right)\\
    g_{25}^{ijkl}&=\frac{\partial^4}{\partial\bm{r}_i\partial\bm{r}_j\partial\bm{\rho}_k\partial\bm{\rho}_l}G_2(\bm{r},\bm{v}_1)J_1\left(\bm{\rho},\frac{E_1 \bm{v}_1}{E_1+m_3}\right)\\
    g_{26}^{ijkl}&= \frac{\partial^4}{\partial\bm{r}_i\partial\bm{r}_j\partial\bm{\rho}_k\partial\bm{\rho}_l}G_2(\bm{r},\bm{v}_1)J_1\left(\bm{\rho},\frac{E_2 \bm{v}_2}{E_2+m_3}\right)\\
    g_{27}&= G_1(\bm{r},\bm{v}_1)G_1\left(\bm{\rho},\frac{E_1 \bm{v}_1}{E_1+m_3}\right)\\
    g_{28}&=G_1(\bm{r},\bm{v}_1)G_1\left(\bm{\rho},\frac{E_2 \bm{v}_2}{E_2+m_3}\right)\\
    g_{29}^{ij}&=- \frac{\partial^2}{\partial\bm{\rho}_i\partial\bm{\rho}_j}G_1(\bm{r},\bm{v}_1)G_2\left(\bm{\rho},\frac{E_1 \bm{v}_1}{E_1+m_3}\right)\\
    g_{210}^{ij}&=- \frac{\partial^2}{\partial\bm{\rho}_i\partial\bm{\rho}_j}G_1(\bm{r},\bm{v}_1)G_2\left(\bm{\rho},\frac{E_2 \bm{v}_2}{E_2+m_3}\right)\\
    g_{211}^{ijkl}&=\frac{\partial^4}{\partial\bm{r}_i\partial\bm{r}_j\partial\bm{\rho}_k\partial\bm{\rho}_l}G_2(\bm{r},\bm{v}_1)G_2\left(\bm{\rho},\frac{E_1 \bm{v}_1}{E_1+m_3}\right)\\
    g_{212}^{ijkl}&=\frac{\partial^4}{\partial\bm{r}_i\partial\bm{r}_j\partial\bm{\rho}_k\partial\bm{\rho}_l}G_2(\bm{r},\bm{v}_1)G_2\left(\bm{\rho},\frac{E_2 \bm{v}_2}{E_2+m_3}\right)\\
    g_{213}^{ij}&= \frac{\partial^2}{\partial\bm{r}_i\partial\bm{r}_j}G_2(\bm{r},\bm{v}_1)G_1\left(\bm{\rho},0\right)
\end{split}
\eeq

\bibliography{biblio}

\end{document}